\documentclass[%
preprint, 
 amsmath,amssymb,
 aps, physrev,
]{revtex4-2}

\usepackage{graphicx}% Include figure files
\usepackage{dcolumn}% Align table columns on decimal point
\usepackage{bm}% bold math
\usepackage{subcaption}
\usepackage{caption}

\usepackage{xcolor}% add hypertext capabilities
\begin{document}
\captionsetup{justification=raggedright, singlelinecheck=false}
\preprint{}

\title{\textbf{Ornstein–Uhlenbeck Process for Horse Race Betting: A Micro–Macro Analysis of Herding and Informed Bettors} 
}% 

\author{Tomoya Sugawara}
\email{tomoya.sugawara723@gmail.com}
\author{Shintaro Mori}%
 \email{Contact author: shintaro.mori@hirosaki-u.ac.jp}
\affiliation{%
Department of Mathematics and Physics,
School of Science and Technology,
Hirosaki University, \\
Bunkyo-cho 3, Hirosaki, Aomori 036-8561, Japan
}%

\date{\today}% It is always \today, today,

\begin{abstract}
We model the time evolution of single-win odds in Japanese horse racing as a stochastic process, deriving an Ornstein–Uhlenbeck (O–U) process by analyzing the probability dynamics of vote shares and the empirical time series of odds movements. Our framework incorporates two types of bettors: herders, who adjust their bets based on current odds, and informed better (fundamentalist), who wager based on a horse’s true winning probability. Using data from 3,450 Japan Racing Association races in 2008, we identify a microscopic probability rule governing individual bets and a mean-reverting macroscopic pattern in odds convergence. This structure parallels financial markets, where traders’ decisions are influenced by market fluctuations, and the interplay between herding and fundamentalist strategies shapes price dynamics. These results highlight the broader applicability of our approach to non-equilibrium financial and betting markets, where mean-reverting dynamics emerge from simple behavioral interactions.
\end{abstract}

\maketitle

%\tableofcontents

\section{\label{sec:introduction}Introduction}

Racetrack betting markets provide a unique environment for studying decision-making under uncertainty. Unlike financial markets, which evolve continuously, racetrack betting is a short-lived, repeated market, making it an ideal system for examining the dynamics of market efficiency and information aggregation \cite{Griffith:1949,Ziemba:1987,Ziemba:2008}. Due to its well-defined probability structure, racetrack betting has attracted researchers from diverse disciplines, including econophysics \cite{Mantegna:1999} and sociophysics \cite{Galam:2008,Galam:2012}, leading to significant insights into decision-making, information aggregation, and collective behavior.

A well-documented phenomenon in racetrack betting is the favorite-longshot bias, where horses with short odds tend to be undervalued, while those with long odds are overvalued \cite{Griffith:1949,Ziemba:2008}. While empirical studies confirm that final odds generally reflect winning probabilities accurately, systematic deviations persist. This raises an important question: what mechanisms drive these deviations, and how does the flow of information shape the evolution of odds?

Previous studies have primarily examined the statistical properties of final odds distributions, revealing power-law behaviors across various racing markets \cite{Park:2001, Ichinomiya:2006, Mori:2010}. However, these models typically assume either a fixed agent type or a constant proportion of agent types, limiting their ability to capture the dynamic nature of information flow. In reality, the proportion of different agent types evolves over time in response to market feedback. To address this limitation, our study introduces a framework in which the proportion of agent types is dynamic, providing deeper insight into the evolution of odds formation.

In this study, we introduce an Ornstein–Uhlenbeck (O–U) process \cite{Uhlenbeck:1930} to model the time evolution of win odds in Japan Racing Association (JRA) horse racing. Our primary contribution is the derivation of a time-dependent O–U process from the probability dynamics of vote shares and the empirical time series of odds movements. By leveraging both theoretical and empirical insights, we establish a framework that captures the evolving structure of the betting market. 

We classify bettors into two groups: herders, who adjust their bets based on current odds, and informed bettors, who incorporate fundamental information about winning probabilities. Our findings reveal that, contrary to previous assumptions \cite{Mori:2010_2}, the proportion of herders decreases over time, leading to a progressively more efficient market. By analyzing a dataset of 3,450 JRA races in 2008, we demonstrate how the interplay between these bettor types shapes market efficiency and price formation.

Our model draws parallels to financial markets, where traders' decisions are influenced by price fluctuations in a feedback loop similar to herding behavior in betting markets \cite{Kanazawa2018}. In foreign exchange and stock markets, the interplay between trend-followers and fundamentalists shapes price dynamics \cite{Lux1995,Bouchaud2002,Alfarano2005}. Our framework follows a similar structure but operates within a closed, short-lived market, providing a controlled environment for studying market efficiency and information aggregation. Specifically, our model is a random walk process in a time-dependent quadratic potential. Compared to the potentials of unbalanced complex kinetics (PUCK) used in foreign exchange markets \cite{MTAKAYASU2006,MIZUNO2007,MTAKAYASU2007,Yamashita2019}, our structure is simpler, 
allowing for a more analytically tractable formulation.

This paper is structured as follows. In Section \ref{sec:model}, we introduce the O–U process to describe the time series of voting fractions, deriving the relationship between the time evolution of mean squared error (MSE) relative to final voting fractions and the drift term of the O–U process. Section \ref{sec:data} analyzes the micro-level dynamics of individual betting probabilities alongside the macro-level evolution of MSE, allowing us to estimate the time-varying proportion of informed bettors. Finally, Section \ref{sec:conclusion} presents the conclusions of this study.

\section{\label{sec:model}Model}
We introduce a voting model in which voters sequentially choose a horse\cite{Mori:2010,Hisakado:2010}. Here, we refer to bettors 
in the win-bet market as voters, as they effectively "vote" for a horse (candidate) among multiple candidates in each race.  

Let $T$ be the total number of voters, and label each voter by their order $t \in \{1,2,\dots,T\}$.  
We denote the final vote share of a horse as $q \in [0,1]$, which represents the objective winning probability of the horse. This probability reflects the efficiency of the horse racing betting market. Horses are distinguished by their final vote shares $q$ in the win-bet market.  

The random variable $X(t, q) \in \{0,1\}$ represents the decision of 
voter $t$ regarding a horse with final vote share $q$.  
If voter $t$ selects the horse, we set $X(t, q) = 1$; otherwise, if the voter selects a different horse, 
then $X(t, q) = 0$.  
The estimated vote share of the horse with vote share $q$ up to voter $t$ is given by:  
\[
Z(t, q) = \frac{1}{t} \sum_{s=1}^{t} X(s, q).
\]

We classify voters into two types:  

1. Informed voters: These voters select a horse with a final vote share of $q$ with probability $q$. Their decisions introduce information into the market. Since their decisions are independent of other voters, they are referred to as independent voters \cite{Hisakado:2010}. The decision function $f_{\text{inf}}(z)$ is given as:
\[
\mathbb{P}(X(t,q)=1|Z(n,q)=z)=f_{\text{inf}}(z)=q.
\]

2. Herding voters: These voters select a horse with a vote share of $z$ with probability $z$. The decision function $f_{\text{herd}}(z)$ is given as:
\[
\mathbb{P}(X(t,q)=1|Z(n,q)=z)=f_{\text{herd}}(z)=z.
\]
In contrast to informed voters, herding voters do not introduce new information into the market. From a game-theoretic perspective, a herding voter follows a Max-Min strategy, minimizing potential losses in the worst-case scenario \cite{Mori:2013}. Since the odds $O$ for a horse with a vote share of $z$ are proportional to $1/z$, the expected return for a herding voter remains independent of $z$.

The behavior of informed voters can be defined with considerable flexibility, and the model described above represents the simplest formulation. If a voter estimates the true winning probability $q$ as $q_{est}$, then their decision function under a given vote share $z$ depends not only on $q_{est}$ and $z$ but also on their individual risk preferences\cite{Manski2006,Wolfers2006}. In general, risk-averse voters may require a larger difference $q_{est} - z$ before betting, while risk-seeking voters may bet even when $q_{est}$ is only slightly greater than $z$. Consequently, the decision function of an informed voter, $f_{\text{inf}}(z)$, is obtained by integrating voter-specific decision thresholds across the distribution of $q_{est}$ and risk preferences. 

As a result, $f_{\text{inf}}(z)$ becomes a smoothly decreasing function of $z$, rather than a simple step function $\theta(q - z)$. Since the market equilibrium closely approximates the true winning probability $q$ and herding voters do not contribute new information, the equilibrium solution $z = f_{\text{inf}}(z)$ satisfies $z = q$. The simplest definition of an informed voter, where $f_{\text{inf}}(z) = q$, ensures that the equilibrium condition $z = f_{\text{inf}}(z)$ is satisfied at $z = q$. While this definition is not strictly decreasing, it captures the fundamental assumption that in an efficient market, the final vote share should align with the true probability $q$.

\begin{figure}[htbp]
\begin{center}
\includegraphics[width=1.0\textwidth]{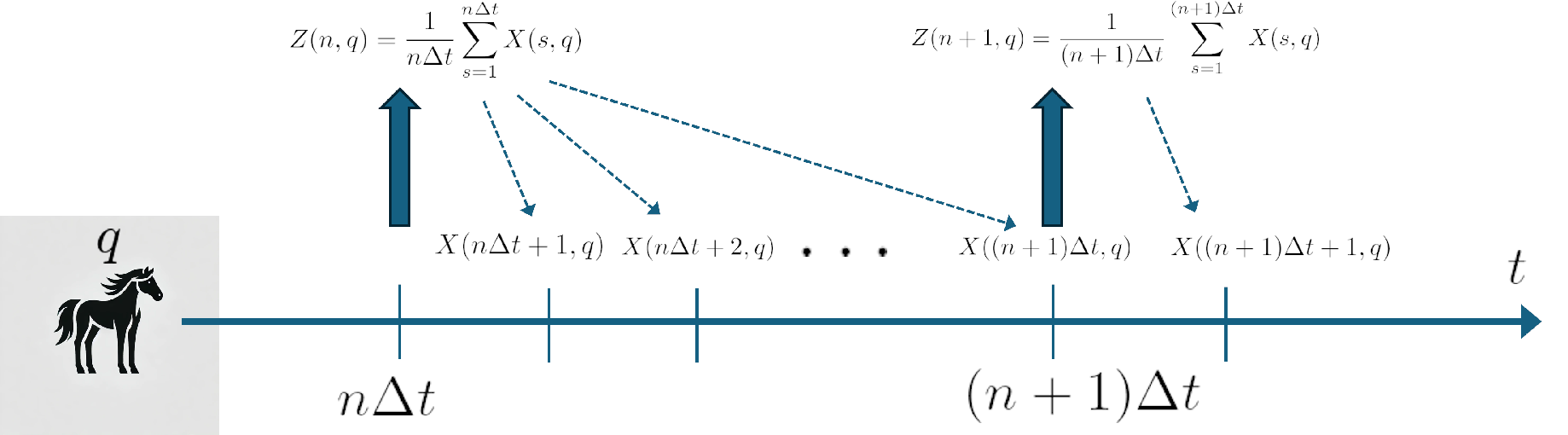}
\end{center}
\caption{Schematic representation of the announcements and the voting process.
Voters sequentially vote for a horse with a winning probability of $q$, and their choices are represented by $X(s,q) \in \{0,1\}$. If voter $s$ chooses the horse, then $X(s,q)=1$; otherwise, $X(s,q)=0$. The $n$-th announcement of the odds is made at $t=n\Delta t$, and the odds are based on the vote share $Z(n,q)$. Voters in the interval $t\in [n\Delta t+1,(n+1)\Delta t]$ observe the odds and make a decision on $X(t,q)\in \{0,1\}$. At $t=(n+1)\Delta t$, the vote share $Z(n+1,q)$ is calculated, 
and the odds are announced.}
\label{fig:model}
\end{figure}

In the JRA win-bet market, the announcement of odds is not real time and there are intervals 
between the anouncements\cite{Mori:2010_2}      .
We denote the interval as $\Delta t$ and the $n$-th announcement is performed after 
voter $t=n\Delta t$. 
Voter $t\in \{n\Delta t+1,\cdots,(n+1)\Delta\},n\in \{1,\cdots,N-1\}$ observes the announcement 
and knows the vote share $Z(n\Delta t,q)$.
Hereafter, we denote the $n$-th announcement of vote share
for horse $q$ as $Z(n,q)$ instead of $Z(n\Delta t,q)$.
Figure \ref{fig:model} illustrates  the announcements and the voting process.
The number of announcement is $N$ and $T=N\Delta t$.
Actually, $\Delta t$ is not constant and depends on the race and $n$.
For more details about the data, please refer to ref.\cite{Mori:2010_2}.

We denote the ratio of informed voter in the $n$-th interval
$[n\Delta t+1,(n+1)\Delta t],n=1,\cdots,N-1$ as $r_{\text{inf}}(n)$.
The voters in the $n$-th interval observes the $n$-th announcement
and knows $Z(n,q)$ of the horse $q$. 
The decision function of voters for the horse $q$
in the $n$-th interval is written as:
\begin{eqnarray}
\mathbb{P}(X(t,q)=1|Z(n,q)=z)&=&f_n(z)=r_{\text{inf}}(n)\cdot f_{\text{inf}}(z)+(1-r_{\text{inf}}(n))\cdot f_{\text{herd}}(z)\nonumber \\
&=&r_{\text{inf}}(n)\cdot q+(1-r_{\text{inf}}(n))\cdot z \label{eq:micro}.
\end{eqnarray}
Equation \eqref{eq:micro} defines the probability that a voter chooses a given horse, 
modeled as a convex combination of herding and informed behaviors. 

We derive the stochastic differential equation (SDE) of $Z(n,q)$.
$Z(n,q)$ is written as,
\[
Z(n,q)=\frac{1}{n\Delta t}\sum_{s=1}^{n\Delta t}
X(s,q).
\]
The difference equation for $Z(n,q)$ is,
\begin{eqnarray}
\Delta Z(n,q)&=&Z(n+1,q)-Z(n,q)=
\frac{1}{(n+1)\Delta t}\sum_{s=1}^{(n+1)\Delta t}X(s,q)
-\frac{1}{n\Delta t}\sum_{s=1}^{n\Delta t}X(s,q) \nonumber \\
&=& \frac{1}{n+1}\left(\frac{1}{\Delta t}\sum_{s=n\Delta t+1}^{(n+1)\Delta t}X(s,q)-Z(n,q)\right)  \nonumber.
\end{eqnarray}
We estimate the conditional expectation and variance of $\Delta Z(n,q)$, 
and derive the stochastic differential equation (SDE) for $Z(n,q)$ following the formalism of Gardiner \cite{Gardiner:2009}.
\begin{equation}
dZ(n,q)=-\frac{r_{\text{inf}}(n)(Z(n,q)-q)}{n+1}dn+\frac{\sqrt{q(1-q)}}{(n+1)\sqrt{\Delta t}}dW(n,q) \label{eq:SDE}.
\end{equation}
Here, $W(n,q)$ is the Wiener process with the increment 
during $[n,n+dn]$ obeys $dW(n,q)=W(n+dn,q)-W(n,q)\sim N(0,dn)$.
This is the time inhomogeneous Ornstein-Uhlenbeck process\cite{Gardiner:2009,Uhlenbeck:1930}.
A detailed derivation of the stochastic differential equation is provided in Appendix A.
As the drift term drive $Z(n,q)$ toward $q$ when $r_{\text{inf}}(n)>0$, which is the mean-reverting of 
the process. In the next section, we see that $r_{\text{inf}}(n)\to 1$ for large $n$ and 
the mean squared error (MSE) between $Z(n,q)$ and $q$ converges to zero.

The potential associated with the process is
\[
U_n(z)=\frac{r_{\text{inf}}(n)}{2(n+1)}(z-q)^2.
\]
Since the potential is time-dependent, the stability of the process depends on 
$r_{\text{inf}}(n)$. As long as $r_{\text{inf}}(n)>0$, the process remains 
stable and the random walker $z$  fluctuates around $q$.

We can integrate the process with the initial condition $Z(1,q)=z(1,q)$
and the solution $Z(n,q)$ is given as,
\begin{eqnarray}
Z(n,q)&=&q+(z(1,q)-q)\exp\left(-\int_{1}^{n}\frac{r_{\text{inf}}(m)}{m+1}dm\right)
\nonumber \\
&+&\int_{1}^n
\exp\left(-\int_{l}^{n}\frac{r_{\text{inf}}(m)}{m+1}dm\right)
\frac{\sqrt{q(1-q)}}{(l+1)\sqrt{\Delta t}}
dW(l,q) \label{eq:sol}.
\end{eqnarray}
The derivation of the solution is given in Appendix B.
As $\mathbb{E}[dW(l,q)]=0$, the expected value of $Z(n,q)$ is,
\[
\mathbb{E}[Z(n,q)]=q+E[z(1,q)-q]\exp\left(-\int_{1}^{n}\frac{r_{\text{inf}}(m)}{m+1}dm\right).
\]

The mean squared error (MSE) of $Z(n,q)$ with respect to $q$ is given by 
\begin{eqnarray}
\mbox{MSE}(Z(n,q),q) &=& \mathbb{E}[(Z(n,q)-q)^2]
= \mbox{MSE}(Z(1,q),q) \exp\left(-2\int_{1}^{n}\frac{r_{\text{inf}}(m)}{m+1}dm\right)\nonumber \\
&+&\int_{1}^{n}
\exp\left(-2\int_{l}^{n}\frac{r_{\text{inf}}(m)}{m+1}dm\right)
\frac{q(1-q)}{(l+1)^2\Delta t}
dl \label{eq:MSE}.
\end{eqnarray}

Below, we summarize the time evolution of the MSE for two cases of 
$r_{\text{inf}}(n)$:  
(1) a constant $r_{\text{inf}}(n)$, and (2) a linearly varying $ r_{\text{inf}}(n)$.

\subsection{Case of Constant $r_{\text{inf}}(n)$}

When $r_{\text{inf}}(n)$ is constant, i.e., $r_{\text{inf}}(n) = r_1$, we obtain  
\begin{eqnarray}
&& \mbox{MSE}(Z(n,q),q) \nonumber \\
&=& \mbox{MSE}(Z(1,q),q) \left(\frac{n+1}{2}\right)^{-2r_{1}}
+\frac{q(1-q)}{\Delta t}(n+1)^{-2r_{1}}\int_{1}^{n}(l+1)^{-2+2r_{1}}dl \nonumber \\
&\simeq & \mbox{MSE}(Z(1,q),q) \left(\frac{n+1}{2}\right)^{-2r_{1}}
+\frac{q(1-q)}{\Delta t}\left\{\begin{array}{cc}
\frac{1}{2r_{1}-1}(n+1)^{-1} & r_{1}>1/2,\\
(n+1)^{-1}\ln \frac{n+1}{2} & r_{1}=1/2,\\
\frac{1}{1-2r_{1}}(n+1)^{-2r_{1}}2^{2r_{1}-1}& r_{1}<1/2.
\end{array}
\right.
\label{eq:MSE1}
\end{eqnarray}

Thus, $\mbox{MSE}(Z(n,q),q)$ decreases following a power law, 
and the exponent changes with $r_{1}$. When $r_{1}>1/2$,
the power law exponent is 1, and MSE decreases as $1/n$ for large $n$.  
When $r_{1}<1/2$,
the power law exponent is $2r_{1}<1$, and the convergence is slower 
than in the case where $r_{1}>1/2$. At $r_{1}=1/2$, a logarithmic correction to the power law $1/n$ occurs\cite{Hisakado:2010,Mori:2015}. 
This transition in asymptotic convergence behavior is known as the super-normal transition.
This transition was first identified by A.~Hod~\cite{Hod:2004} in a model describing long-term correlations in the stock market and DNA sequences.  
Huillet also discussed a similar phase transition in the power-law behavior of the Friedman-Pólya process~\cite{Huillet:2008}.

It should be noted that $\mbox{MSE}(Z(n,q),q)$ consists of two terms, and the observed power-law behavior depends on which term dominates. For sufficiently large $n$, the term with the smaller exponent determines the asymptotic behavior. However, when $n$ is not large enough, the observed exponent is influenced by the relative magnitude of the two terms, making it difficult to detect the true asymptotic exponent.

\subsection{Case of Linearly Varying $r_{\text{inf}}(n)$}
Before analyzing the implications of a linearly varying $r_{\text{inf}}(n)$, we briefly clarify the motivation for adopting this functional form. The linear dependence is not assumed a priori, but is instead guided by empirical findings. As shown in Section~\ref{sec:data}, the estimated proportion of informed voters $r_{\text{inf}}(n)$ increases approximately linearly with the normalized time $n$ in the Japanese horse racing data. From a theoretical perspective, this linear form introduces a simple yet effective nonstationarity into the O–U framework, allowing us to analytically describe the crossover from power-law convergence of the mean squared error (MSE) to an accelerated exponential decay.

Let $\Delta r= r_{\text{inf}}(N) - r_{\text{inf}}(1)$ be the total change in $r_{\text{inf}}(n)$.  
We express $r_{\text{inf}}(n)$ as  
\[
r_{\text{inf}}(n) = r_{1} + \frac{n-1}{N-1} \Delta r.
\]
From the results for the constant $r_{\text{inf}}(n) = r_1$ case, we intuitively expect that when $r_{\text{inf}}(n)$ changes over time, the power-law exponent governing the convergence of MSE should also evolve over time. If $\Delta r> 0$, the exponent increases over time, leading to a steeper decay in MSE. Consequently, in a double logarithmic plot of MSE, the absolute value of the slope is expected to increase over time.

The MSE of $Z(n,q)$ with respect to $q$ is then estimated as  
\begin{eqnarray}
&&\mbox{MSE}(Z(n,q),q) \simeq  \left(\frac{n+1}{2}\right)^{-2r_{1}+4\frac{\Delta r}{N-1}}e^{-2\frac{n-1}{N-1}\Delta r}
\nonumber \\
&\times& \left(\mbox{MSE}(Z(1,q),q)+\frac{q(1-q)}{\Delta t}
2^{-2r_{1}+4\frac{\Delta r}{N-1}}\int_{1}^{n}(l+1)^{-2+2r_{1}-4\frac{\Delta r}{N-1}}e^{2\frac{l-1}{N-1}\Delta r}dl
\right) \label{eq:MSE2}.
\end{eqnarray}
Since $\Delta r \ll N$, we can neglect $\Delta r/(N-1)$.
In contrast to the case where $r_{\text{inf}}$ is constant, 
the integral in the bracket converges in the limit $n \gg 1$ when $\Delta r > 0$.  
The asymptotic behavior of $\mbox{MSE}(Z(n,q),q)$ is then governed by 
the prefactor before the bracket.
In addition, when the first term $\mbox{MSE}(Z(1,q),q)$ in the bracket is much larger 
than the second term, we obtain
\begin{equation}
\mbox{MSE}(Z(n,q),q) \simeq \left(\frac{n+1}{2}\right)^{-2r_{1}+4\frac{\Delta r}{N-1}}e^{-2\frac{n-1}{N-1}\Delta r} \times \mbox{MSE}(Z(1,q),q).
\label{eq:MSE3}
\end{equation}
For small $n$, $\mbox{MSE}$ decreases as a power law, with the power-law exponent 
given by $2r_{1}$. When $\Delta r> 0$,
one also observes an exponential decay for $n > n_c=(N-1)/(2\Delta r)+1$.
When $\Delta r < 0$, the power-law convergence is  hindered by exponential growth.

In the next section, we analyze the microscopic probabilistic rules governing voter behavior in  
Eq.\eqref{eq:micro} and the macroscopic convergence behavior of the MSE.  

\section{\label{sec:data}Data Analysis}

We analyze the win bet data from JRA races in 2008.  
A win bet is a wager placed on the horse that finishes first in a race.  
A total of 3,450 races were held that year, which we index as $r = 1, \dots, R = 3450$.  
The final public win pool (i.e., the total number of votes placed on win bets)  
for race $r$ is denoted as $T[r]$.  
The values of $T[r]$ range from $5.9\times 10^4$ to $1.46 \times 10^7$,  
with an average of a approximately $3.0 \times 10^5$.  
Each race $r$ had between $H[r] \in [7,18]$ participating horses.  
The total number of horses included in the dataset is 50180.  
The number of winning horses is 3453 (with 3 cases of ties).  

We denote by $I[r]$ the number of public announcements made  
regarding the temporal odds and the total number of votes 
during race $r$.  
The values of $I[r]$ range from $14$ to $401$,  
with an average of approximately 80 announcements before the race started.  
The temporal odds of the $h$-th horse in race $r$ at the $i$-th announcement  
are denoted as $O[r,i,h]$, while the total number of votes at that time  
is denoted as $t[r,i]$.  
By definition, the final total number of votes satisfies 
$t[r, I[r]] = T[r]$.

\begin{table}[htbp]
\centering
\caption{A sample of time series of odds and pool for a race that starts at $13:00$. 
The race consists of $H[r]=10$ horses and $I[r]=53$ announcements. 
We show data for the first three horses ($1 \leq h \leq 3$) and the last horse ($h=10$).}
\label{tab:TS}
\resizebox{0.8\textwidth}{!}{
\begin{tabular}{r r r r r r c r}
\hline
$i$ & Time to Post [min] & $t[r,i]$ & $O[r,i,1]$ & $O[r,i,2]$ & $O[r,i,3]$ & $\cdots$ & $O[r,i,10]$ \\
\hline
1  & 358  &       1  &  0.0  &  0.0  &  0.0  & $\cdots$ & 0.0 \\  
2  & 351  &     169  &  1.6  & 33.3  &  7.9  & $\cdots$ & 33.3 \\   
3  & 343  &     314  &  1.8  & 11.3  &  8.0  & $\cdots$ &  6.7 \\  
4  & 336  &     812  &  2.9  & 17.8  & 14.6  & $\cdots$ &  9.9 \\  
5  & 329  &    1400  &  3.3  &  8.6  & 10.6  & $\cdots$ &  5.7 \\  
6  & 322  &    1587  &  2.7  &  9.2  & 11.3  & $\cdots$ &  6.1 \\  
$\vdots$ & $\vdots$ & $\vdots$ & $\vdots$ & $\vdots$ & $\vdots$ & $\vdots$ & $\vdots$ \\
51 &  10  &   80064  &  2.4  &  6.4  & 13.4  & $\cdots$ &  8.0 \\  
52 &   4  &  148289  &  2.4  &  4.9  & 16.1  & $\cdots$ &  8.2 \\  
53 &  -2  &  211653  &  2.4  &  5.3  & 17.0  & $\cdots$ &  7.9 \\  
\hline
\end{tabular}} 
\end{table}

\subsection{Data Preprocessing}
Here, we describe the data preprocessing steps in detail.
\begin{enumerate}
\item \textbf{New time variable $n$}
The natural time variable in the betting process might be the number of votes $t[r,i]$.  
However, since the total number of votes $T[r]$ varies significantly across races,  
we introduce a normalized time variable $n$ to ensure comparability:  
\[
n[r,i] = 100 \times \frac{t[r,i]}{T[r]}.
\]
We set the initial time as $n[r,0] = 0$ and define the normalized time range as  
\[
n[r,i] \in [0,100], \quad i = 0, \dots, I[r].
\]
While $n$ serves as a standardized measure of time progression in the betting process,  
it does not correspond to evenly spaced real-time intervals before the race starts.  
For instance, as shown in Fig.~\ref{fig:n_vs_T}, at $n=1$, the remaining time until 
the race start is approximately 480 minutes,  
whereas at $n=50$, it is reduced to about 20 minutes.  
This indicates that the interval between consecutive values of $n$ 
shrinks as the race approaches, suggesting that bettors' reactions to odds announcements might differ significantly depending on the time remaining before the race.

\begin{figure}[htbp]
    \begin{center}
    \includegraphics[width=0.8\textwidth]{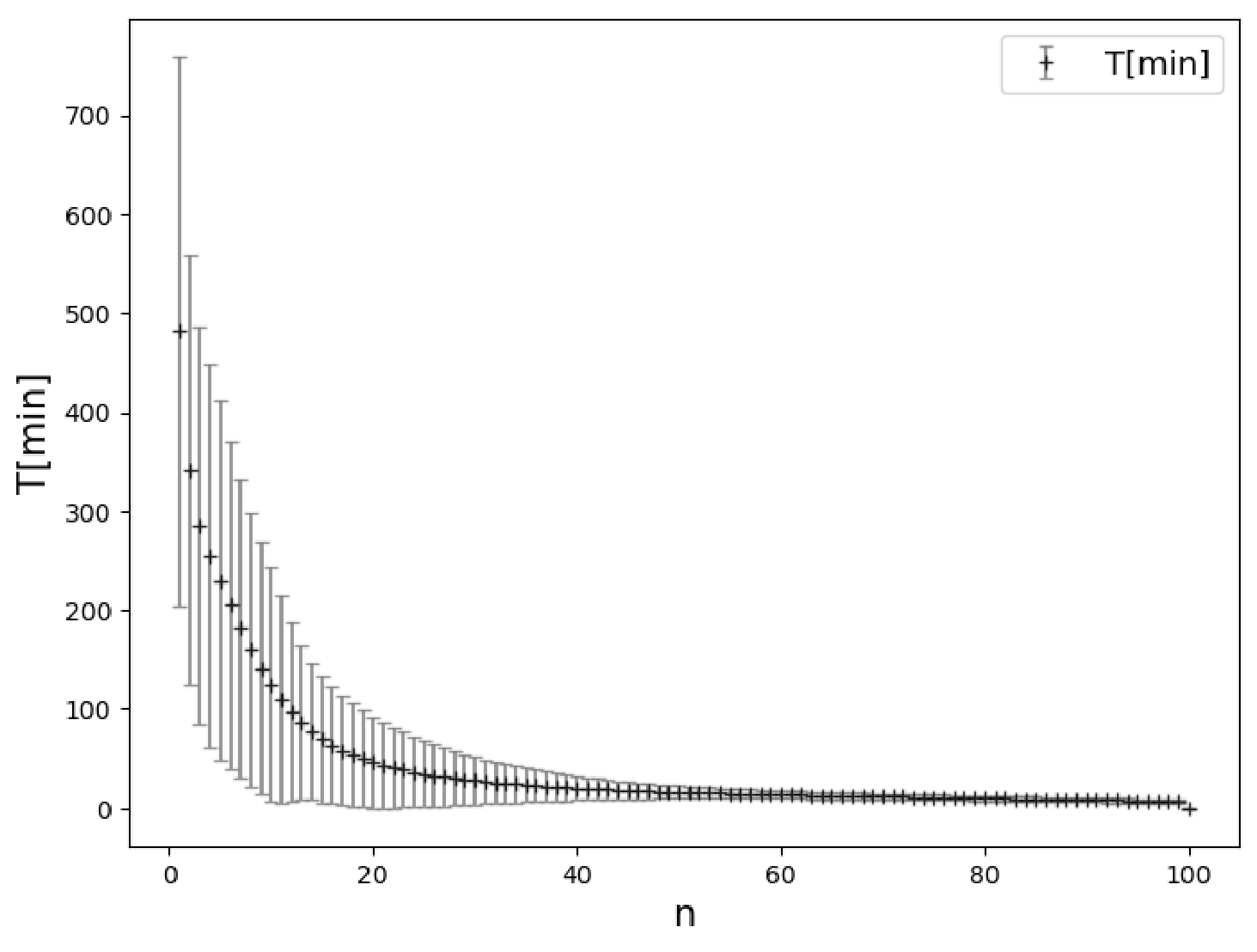}
    \end{center}
    \caption{Relationship between the normalized time variable $n$ and the remaining time until post time.
    The error bars represent the standard errors of the mean remaining time before post, for each normalized time point $n$.
    The time intervals between successive values of $n$ are not uniform, with a significant acceleration in betting activity as the race approaches.}
    \label{fig:n_vs_T}
\end{figure}

\item \textbf{Continuous-time odds $O(r,h,n)$}

We interpolate the odds between consecutive announcements to obtain  
continuous-time odds, denoted as $O(r,h,n)$.  
For $n \in (n[r,i-1], n[r,i])$, we estimate $O(r,h,n)$ using linear interpolation:
\[
O(r,h,n) = \frac{n[r,i] - n}{n[r,i] - n[r,i-1]} O[r,h,i-1] 
+ \frac{n - n[r,i-1]}{n[r,i] - n[r,i-1]} O[r,h,i].
\]

\item \textbf{Continuous-time vote share $Z(r,h,n)$}

In general, the vote share $Z$ is converted into odds $O$ using the following formula,  
as defined by JRA:
\[
O = \max \left(1.1, \frac{0.788}{Z} \right).
\]
After calculating the odds, any decimal places beyond the second digit are truncated.  
Due to this truncation, there is some uncertainty in the conversion from odds $O$ to vote shares $Z$.  
To mitigate this issue, we adjust the odds by adding $0.05$.  
The vote share is then computed as  
\[
Z(r,h,n) = \frac{C}{O(r,h,n)+0.05}.
\]
Here, the normalization constant $C$ is chosen such that  
the sum of $Z(r,h,n)$ over all $H[r]$ horses in race $r$ equals 1:
\[
\sum_{h=1}^{H[r]} Z(r,h,n) = 1.
\]
We denote the final vote share of horse $h$ in race $r$ as $q(r,h)=Z(r,h,100)$.
The Python code for the subsequent analysis is available on GitHub~\cite{DA}.
\end{enumerate}

\subsection{Microscopic Analysis: Voting Rule}
We analyze the microscopic probability rule described in Eq.\eqref{eq:micro} 
using Eq.\eqref{eq:micro2}.  
We estimate $Z(r,h,n) - q(r,h)$ as $Z(n,q) - q$  
and $(n+1)(Z(r,h,n+1) - Z(r,h,n))$ as $(n+1) \Delta Z(n,q)$.  

\begin{figure}[htbp]
    \centering
    \begin{subfigure}{0.45\textwidth}
        \centering
        \includegraphics[width=\textwidth]{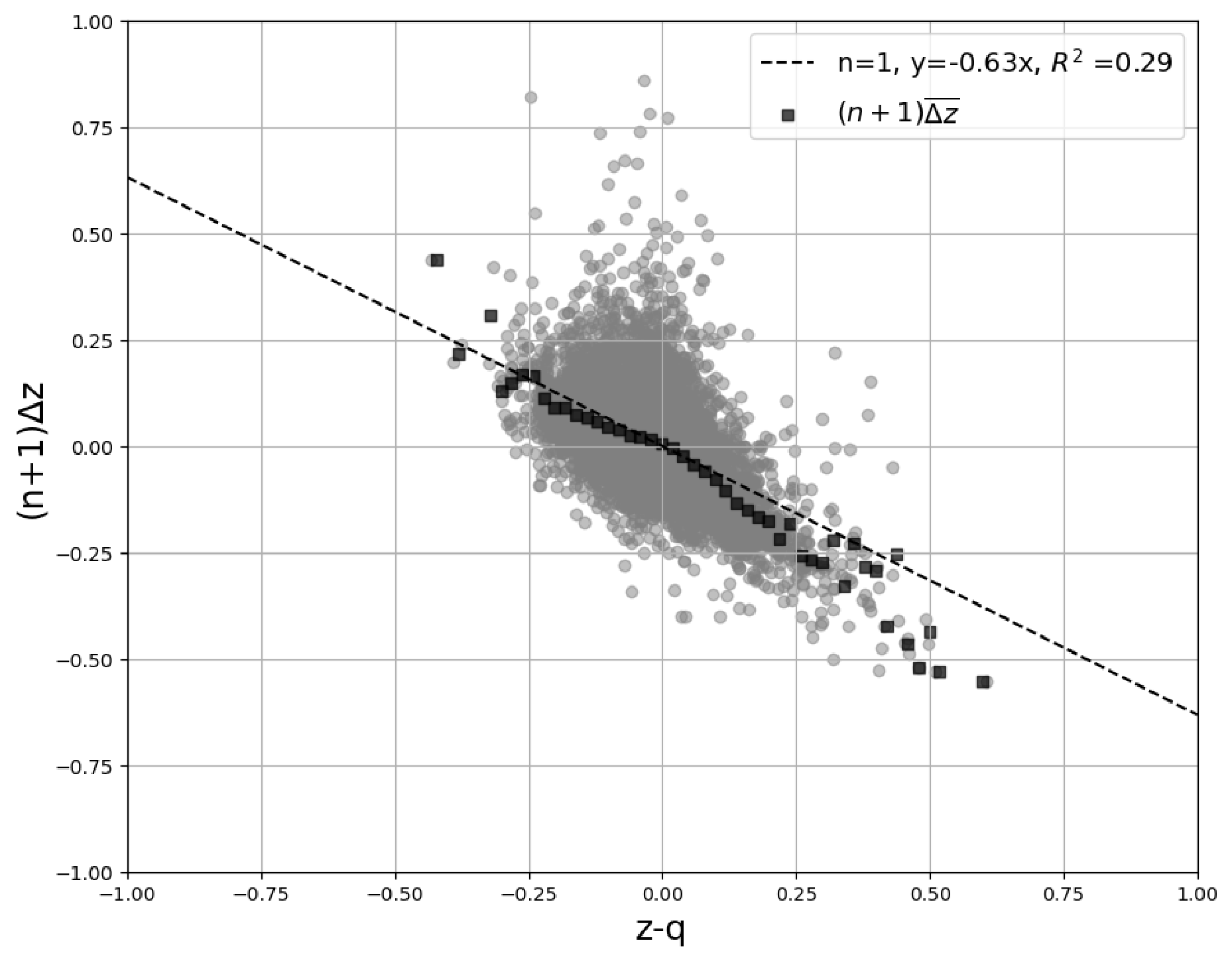}
        \caption{$n=1$}
    \end{subfigure}
    \begin{subfigure}{0.45\textwidth}
        \centering
        \includegraphics[width=\textwidth]{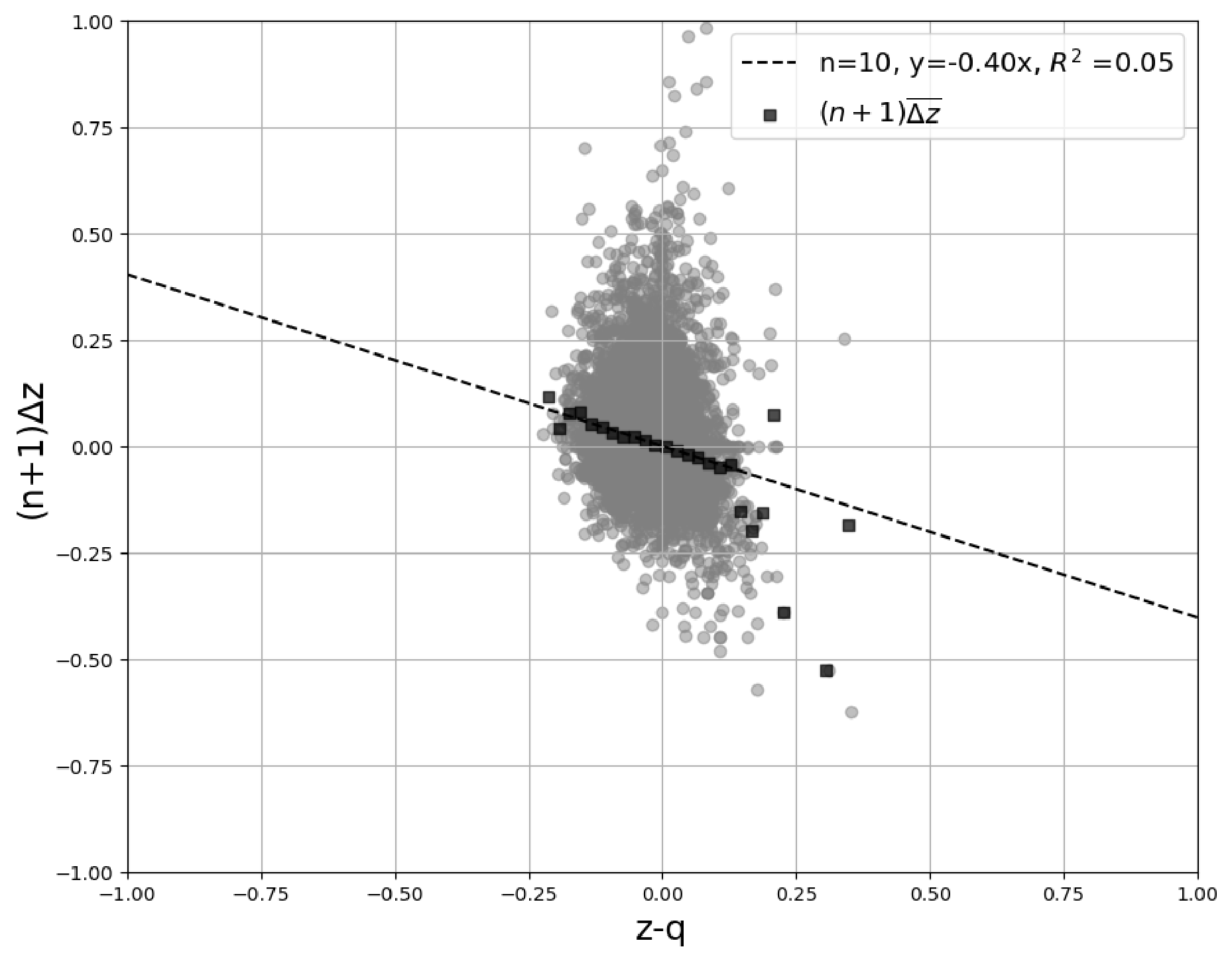}
        \caption{$n=10$}
    \end{subfigure}

    \begin{subfigure}{0.45\textwidth}
        \centering
        \includegraphics[width=\textwidth]{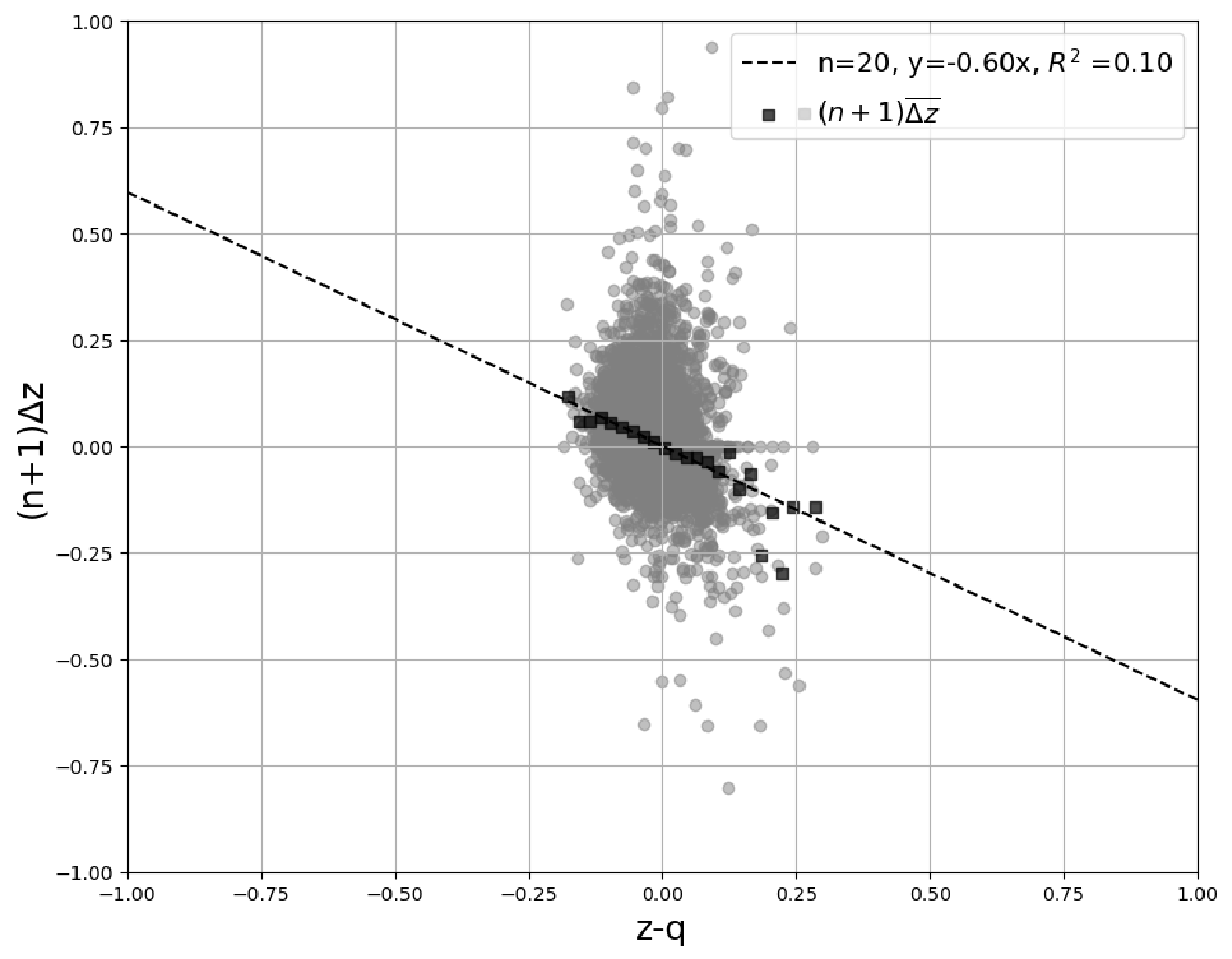}
        \caption{$n=20$}
    \end{subfigure}
    \begin{subfigure}{0.45\textwidth}
        \centering
        \includegraphics[width=\textwidth]{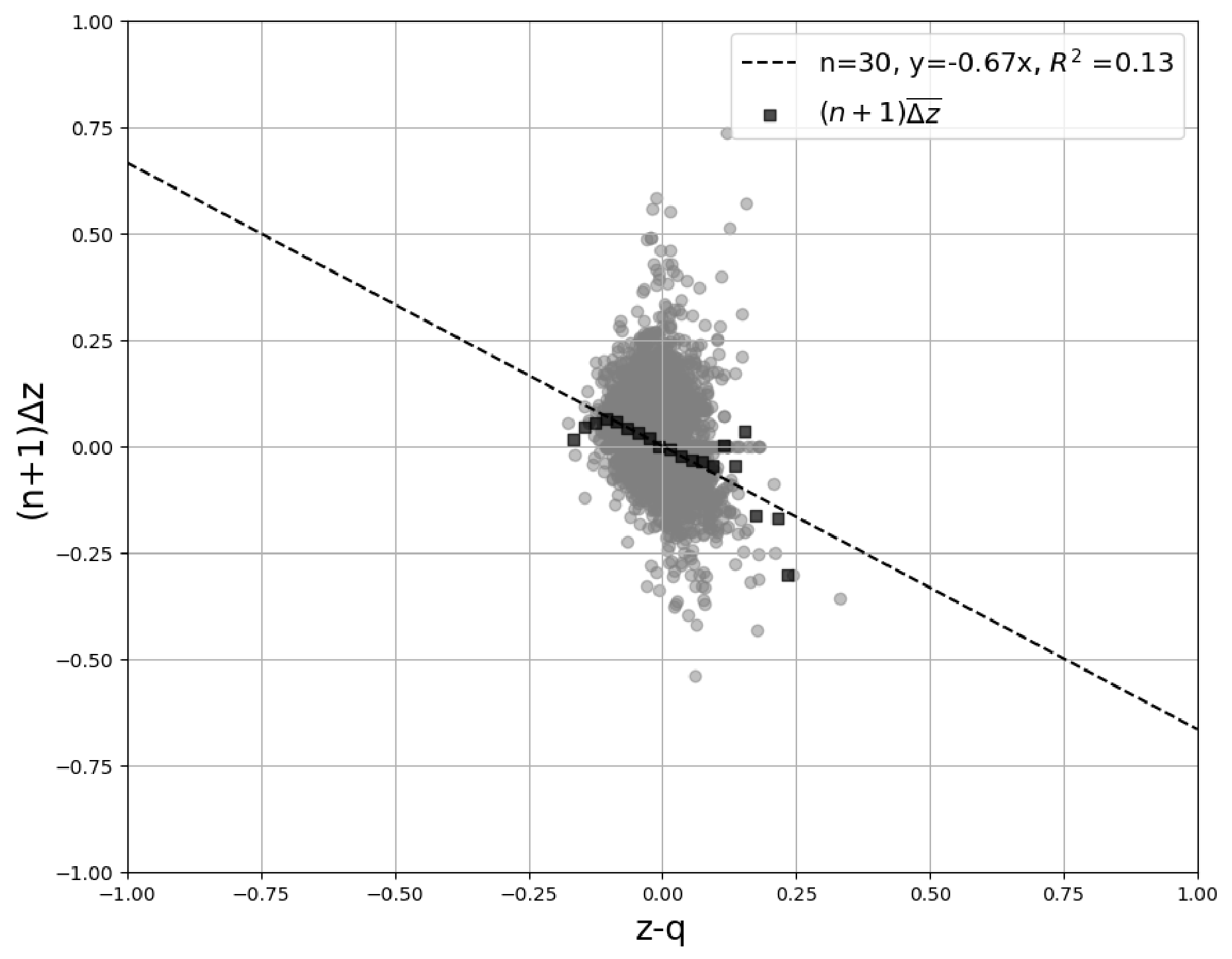}
        \caption{$n=30$}
    \end{subfigure}

    \begin{subfigure}{0.45\textwidth}
        \centering
        \includegraphics[width=\textwidth]{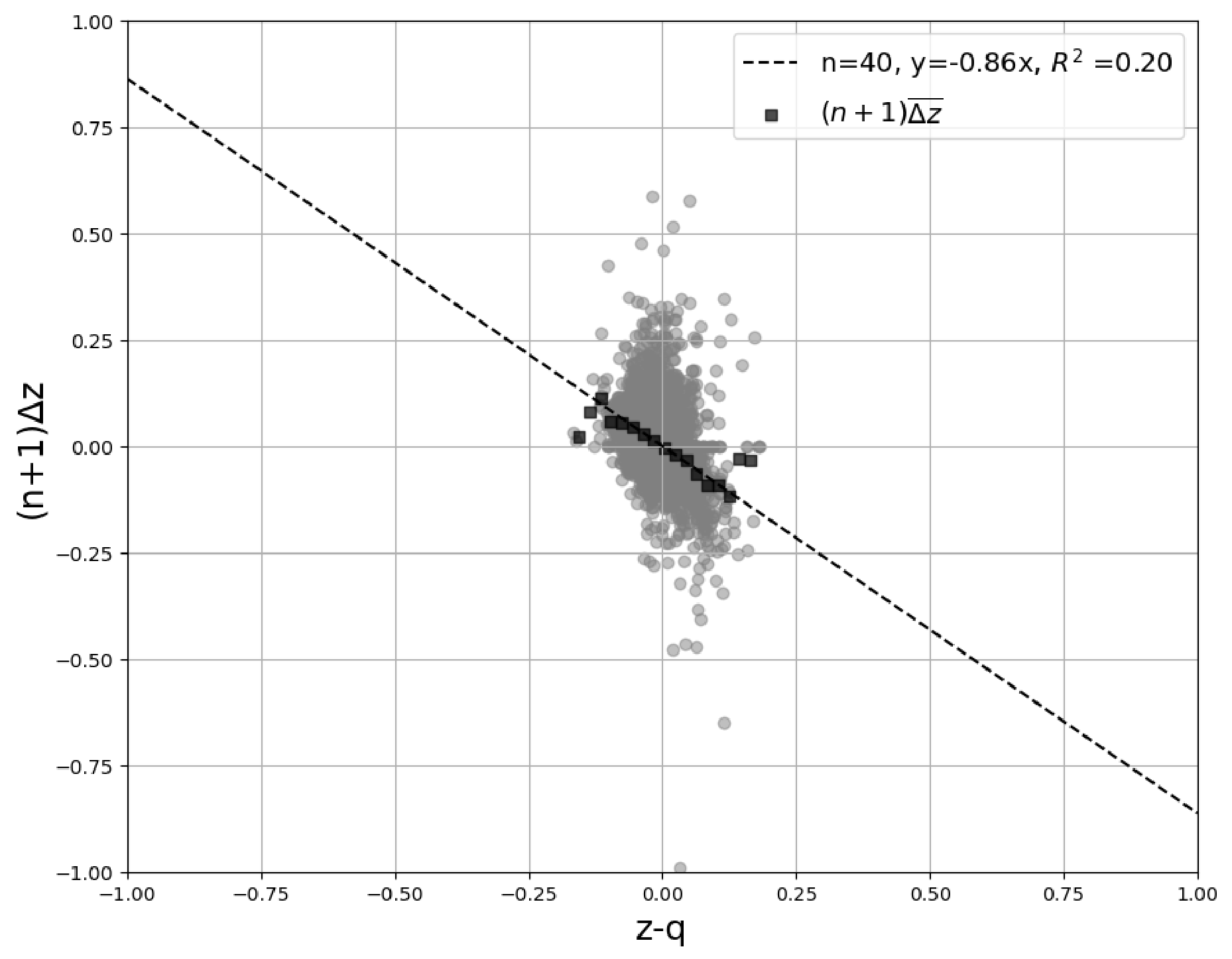}
        \caption{$n=40$}
    \end{subfigure}
    \begin{subfigure}{0.45\textwidth}
        \centering
        \includegraphics[width=\textwidth]{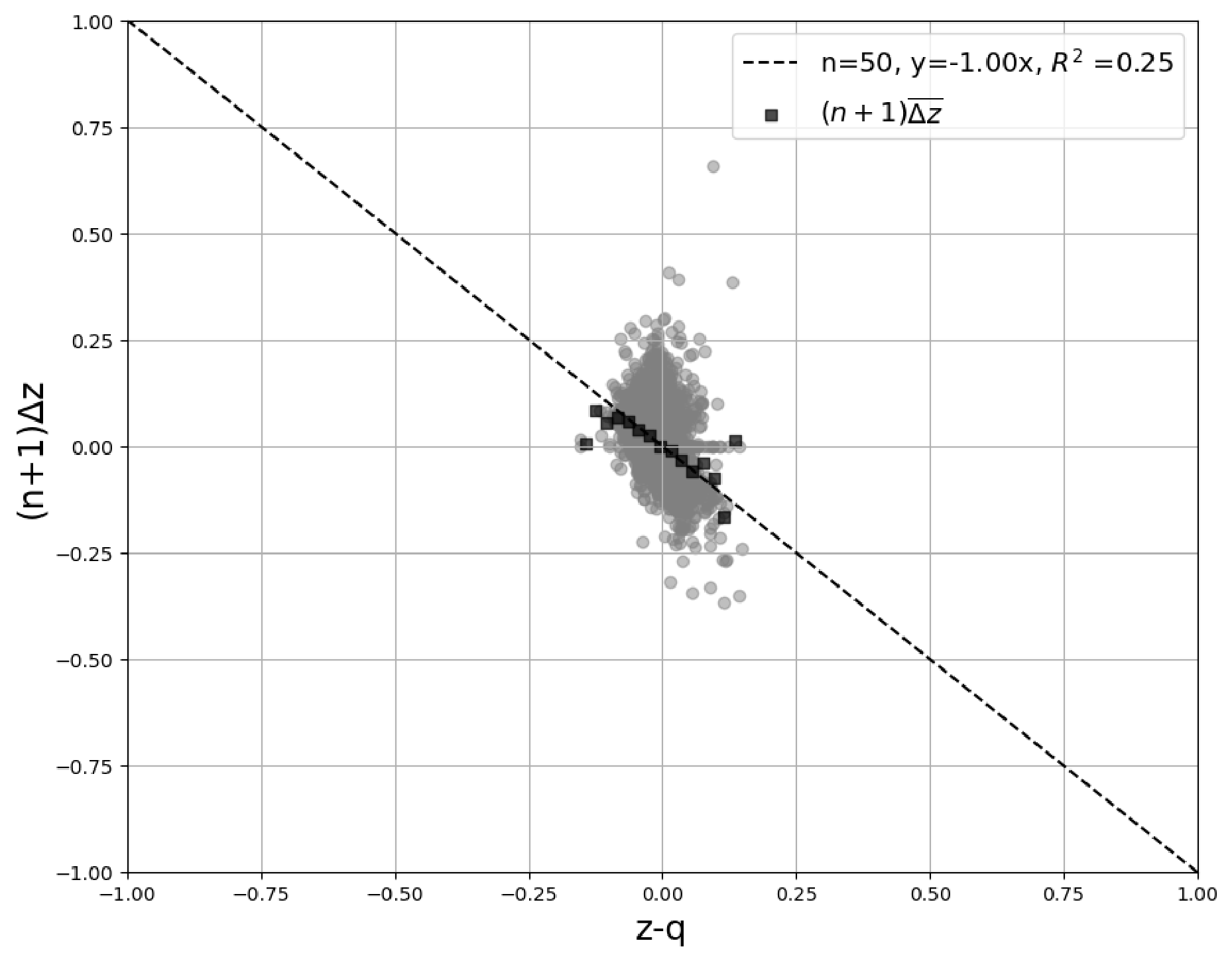}
        \caption{$n=50$}
    \end{subfigure}
    \caption{Scatter plots of $(n+1)(Z(r,h,n+1) - Z(r,h,n))$ versus $Z(r,h,n) - q(r,h)$ 
    for $n \in \{1,10,20,30,40,50\}$ (gray circles).
    The black circles show the mean values of $(n+1)(Z(r,h,n+1) - Z(r,h,n))$
    for each $Z(r,h,n) - q(r,h)$. 
    The broken lines indicate the results of linear regression with zero intercepts.}
    \label{fig:micro}
\end{figure}

Figure \ref{fig:micro} presents scatter plots of  
$(n+1)(Z(r,h,n+1) - Z(r,h,n))$ versus $Z(r,h,n) - q(r,h)$  
for $n \in \{1,10,20,30,40,50\}$.  
The black circles represent the mean values of $(n+1)(Z(r,h,n+1) - Z(r,h,n))$  
for each $Z(r,h,n) - q(r,h)$. The mean of $(n+1)(Z(r,h,n+1) - Z(r,h,n))$  
exhibits a linear dependence on $Z(r,h,n) - q(r,h)$,  
suggesting that Eq. \eqref{eq:micro} is a good model for describing this dependence.  
The broken lines represent the results of linear regression  
with a zero intercept, performed using the least squares method:
\begin{equation}
(n+1)(Z(r,h,n+1)-Z(r,h,n)) = -r_{\text{inf}}(Z(r,h,n)-q(r,h)).
\end{equation}  
From Eq. \eqref{eq:micro2}, the regression coefficient estimates $-r_{\text{inf}}$.  
We only present results for $n \leq 50$.  
For $n > 50$, since $Z(r,h,n)$ nearly coincides with $q(r,h)$,  
it becomes difficult to estimate the regression coefficients.  
The regression coefficient $-r_{\text{inf}}$ 
decreases with $n$ and approaches $-1$ at $n=50$.  

\begin{figure}[htbp]
    \begin{center}
    \includegraphics[width=0.6\textwidth]{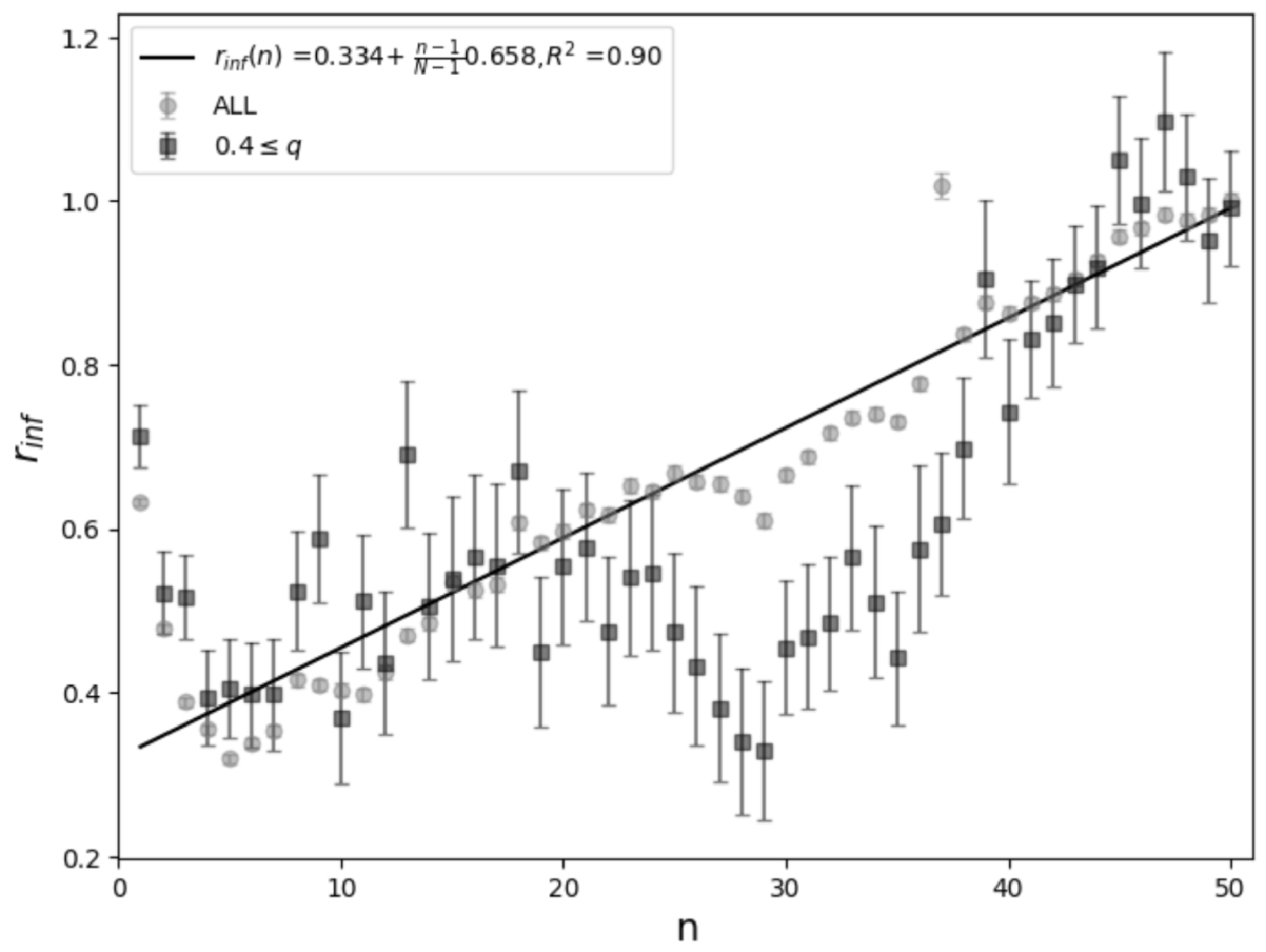}
    \end{center}
    \caption{Plot of $r_{\text{inf}}$ as a function of $n$. 
    Gray circles represent all horses, and black squares indicate horses with $q\ge 0.4$. 
    The solid black line shows the linear regression result for all horses: 
    $r_{\text{inf}}(n) = 0.334 + \frac{n-1}{50-1} \cdot 0.658$, with a coefficient of determination $R^2 = 0.90$.}
\label{fig:r_inf}
\end{figure}

The low $R^2$ values observed in the regression models can be attributed to two key sources of heterogeneity in the data.
First, the regression is performed over all horses without stratifying by their final winning probability $q$. As shown in Fig.~\ref{fig:r_inf}, the inferred ratio of informed bettors $r_{\text{inf}}(n)$ depends significantly on $q$. This 
unaccounted-for heterogeneity introduces noise into the regression, thereby reducing the explanatory power of a single 
linear model and resulting in lower $R^2$ values. Second, the normalized time variable $n$ used in our analysis aggregates betting behavior across races, but it does not correspond to consistent real-time intervals. As illustrated in Fig.~\ref{fig:n_vs_T}, the standard error of the remaining time until post 
time at a given $n$ is large. For instance, while $n = 10$ may correspond to just a few minutes before post time in one race, it may occur 200 minutes prior in another. This inconsistency means that the same value of $n$ may reflect very different stages of decision-making in different races. Consequently, bettor behavior at a given $n$ is not homogeneous across races, and this further contributes to the low $R^2$ values observed in the aggregated regression analysis.

Figure \ref{fig:r_inf} presents a plot of $r_{\text{int}}$ as a function of $n$.
The gray circles represent the results for all horses, while the black boxes 
indicate the results for horses with $q>0.4$.
As one can clearly see, $r_{\text{inf}}$ increases linearly with $n$ in the case of all horses.  
This increase in $r_{\text{inf}}(n)$ can be interpreted behaviorally: as the race approaches and the odds stabilize,
bettors become more confident in their assessments. Consequently, their behavior shifts from imitation to more
informed decision-making based on subjective probability judgments. This reflects a natural transition from herding to 
rational estimation as uncertainty diminishes.

To assess the validity of the linear assumption for $r_{\text{inf}}(n)$, we performed a linear 
regression for the full dataset. \color{black}
We assume the linear dependence of $r_{\text{inf}}$ as
\[
r_{\text{inf}}(n)=r_{1}+\frac{n-1}{50-1}\Delta r,
\]
and obtained a high coefficient of determination $(R^2=0.90)$, confirming the appropriateness of 
the linear model in this case.
We estimate the regression coefficients $r_1$ and $\Delta r$ as
$r_1=0.334$ and $\Delta r=0.658$. At $n=50$, $r_{\text{inf}}$ becomes 
$0.334+0.658=0.992$,
and the ratio of herding voters is $0.008$.

We also investigate the dependence of $r_{\text{inf}}(n)$ on $q$.
We stratify the horses into a group based on their $q$ values, with $q \geq 0.4$.
Approximately 1.6\% of the horses belong to this category.
$r_{\text{inf}}$ fluctuates around 0.4 for $n\leq 30$
and increases to 1 for $n>30$. 
Thus no linear fit or $R^2$ value is reported for this subset.
The difference from the results in the case of all horses 
is interesting. The ratio of herders remains high for long duration of the voting process.
This results in slower convergence of $Z(n,q)$ to $q$ and a favorite bias for horses with short odds. 
However, the data for $q>0.4$ is limited and one need more data
to determine the dependence of $r_{\text{n}}$ on $n$.

\subsection{Macroscopic Analysis: Convergence of MSE}

We analyze the dependence of the MSE on $n$ by evaluating the mean value of  
$(Z(r,h,n) - q(r,h))^2$ as an estimator of  
$\mbox{MSE}(Z(n,q),q)$.  
Since $Z(r,h,n) = q(r,h)$ at $n=100$, it follows trivially that $\mbox{MSE}(Z(100,q),q) = 0$.  
Thus, we cannot examine the convergence behavior for large $n \approx 100$.  
Instead, we investigate the convergence behavior for $n \leq 50$.  

As $r_{\text{inf}}(n)$ increases linearly with $n$ in the case of all horses,  
we can apply the theoretical results on the convergence behavior of $Z(n,q)$  
to $q$, as given in Eqs.~\eqref{eq:MSE2} and \eqref{eq:MSE3}.  
Furthermore, since $\Delta t \simeq 3\times 10^3$ and $N=10^2 \gg \Delta r$,  
the numerical estimate of the second integral term in Eq.~\eqref{eq:MSE2}  
is much smaller than the first term, $\mbox{MSE}(Z(1,q),q)$.  
Thus, we adopt the approximate form of Eq.~\eqref{eq:MSE3},  
and the logarithm of the ratio of $\mbox{MSE}(Z(n,q),q)$ to $\mbox{MSE}(Z(1,q),q)$ satisfies  
\[
\ln \left(\frac{\mbox{MSE}(Z(n,q),q)}{\mbox{MSE}(Z(1,q),q)}\right) \simeq  
\left( {-2r_{1}+4\frac{\Delta r}{N-1} }\right) \ln \left(\frac{n+1}{2}\right) - 2\left(\frac{n-1}{N-1}\right)\Delta r.
\]
For small $n$, we can neglect the second term since $n \ll N$,  
and the absolute value of the slope of $\ln (\mbox{MSE}(Z(n,q),q)/\mbox{MSE}(Z(1,q),q))$  
is given by $2r_1$. As $n$ increases, the second term contributes,  
leading to an increase in the slope.  

\begin{figure}[htbp]
    \centering
    \begin{subfigure}{0.45\textwidth}
        \centering
        \includegraphics[width=\textwidth]{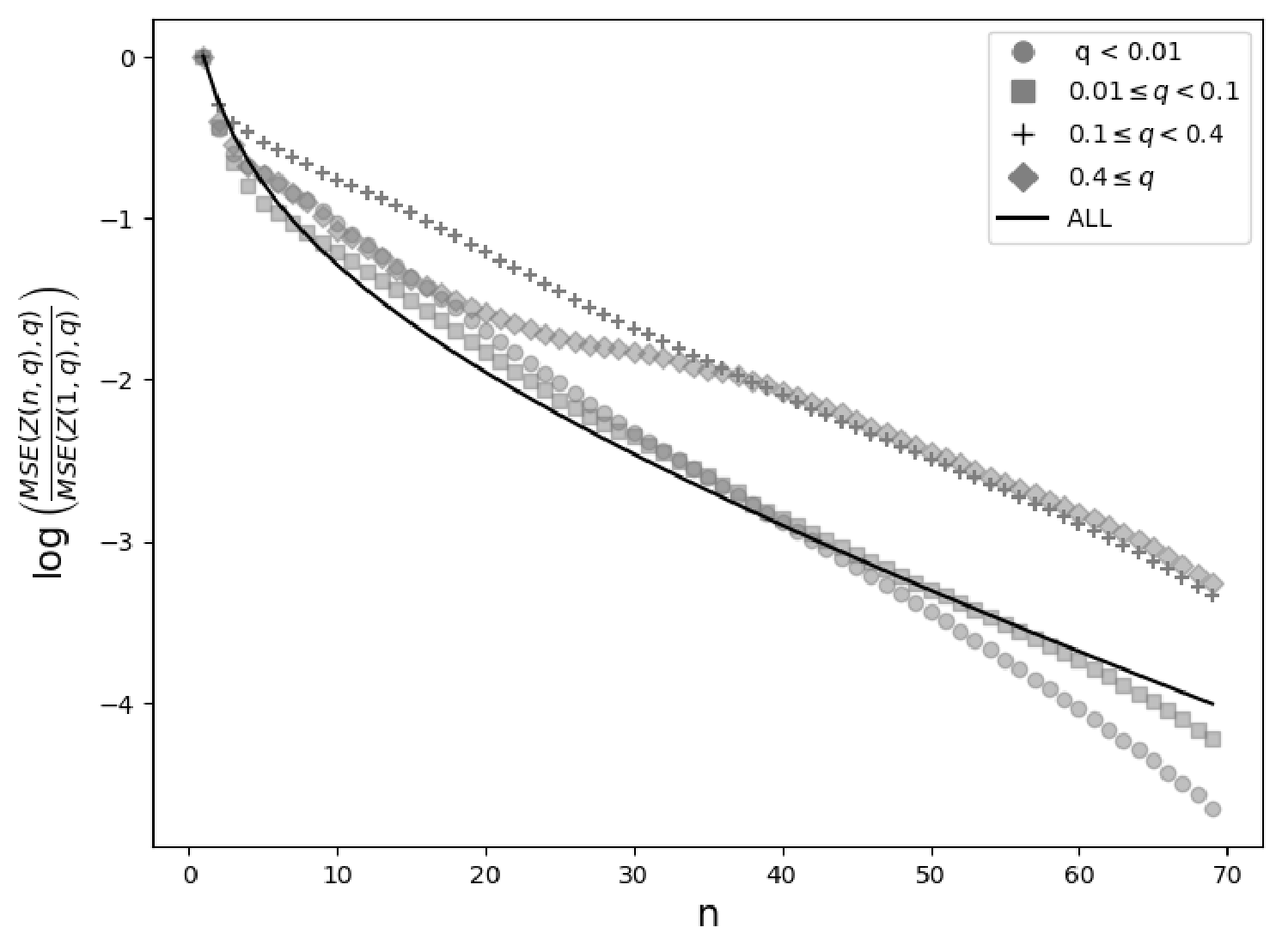}
    \end{subfigure}
    \begin{subfigure}{0.45\textwidth}
        \centering
        \includegraphics[width=\textwidth]{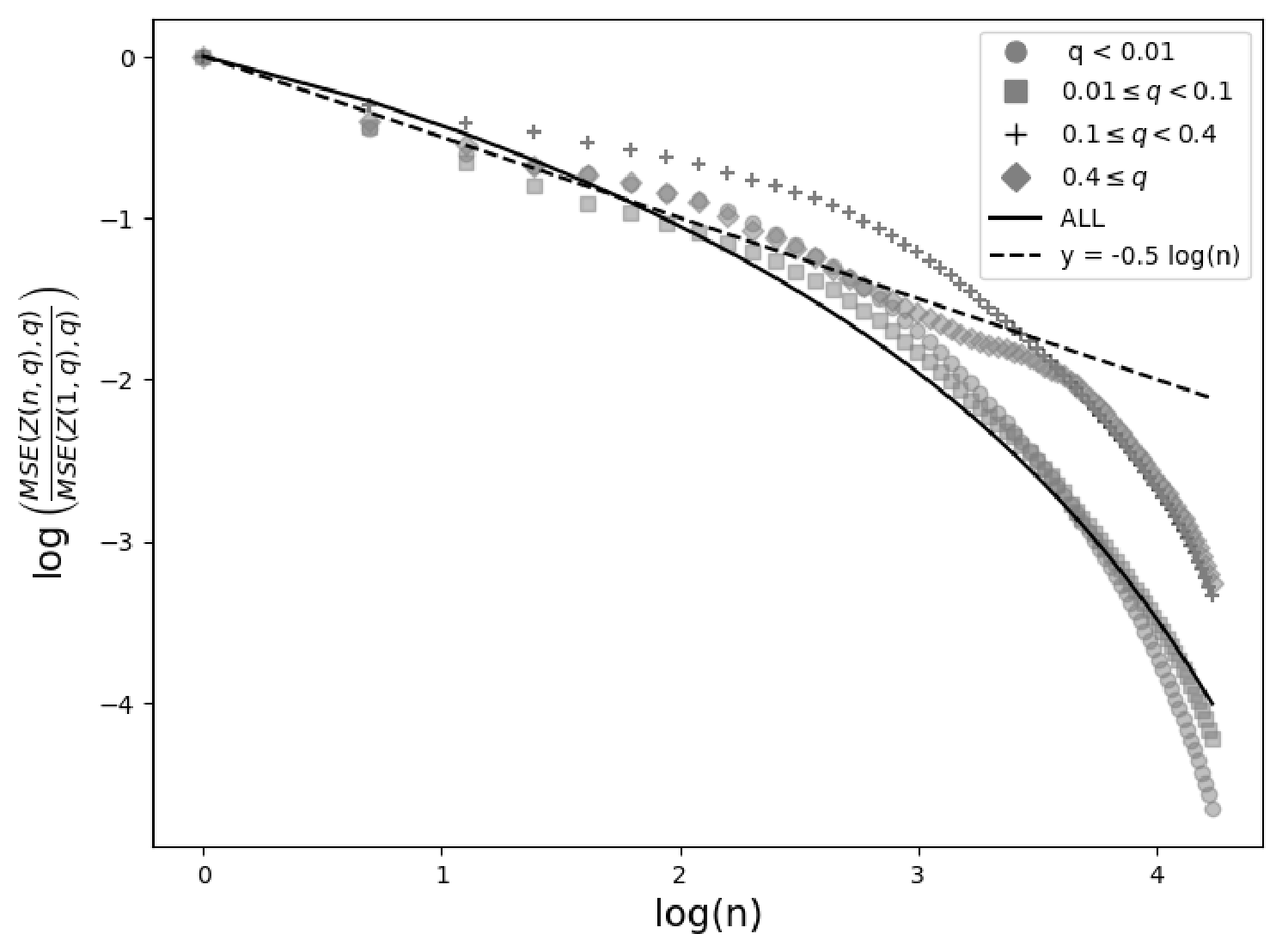}
    \end{subfigure}
    \caption{Plots of the logarithm of the mean values of $(Z(r,h,n) - q(r,h))^2$  
    normalized by the value at $n=1$ as a function of $n$ (Left) and $\ln(n)$ (Right).  
    The results for $q < 0.01$ (circles), $0.01 \leq q < 0.1$ (squares),  
    $0.1 \leq q < 0.4$ (crosses), and $q \geq 0.4$ (diamonds).  
    The solid line represents the theoretical results from Eq.~\eqref{eq:MSE2}.  
    }
    \label{fig:MSE}
\end{figure}

Figure~\ref{fig:MSE} presents the logarithm of the mean values  
of $(Z(r,h,n) - q(r,h))^2$ normalized by the value at $n=1$,  
plotted as a function of $n$ (Left) and $\ln(n)$ (Right).  
We stratify the data into four cases: $q < 0.01$, $0.01 \leq q < 0.1$,  
$0.1 \leq q < 0.4$, and $q \geq 0.4$,  
with their respective proportions being about 25.6\%, 52.2\%, 20.6\%, and 1.6\%.  
Both plots start at $(1,0)$ and decrease as functions of $n$.  
The theoretical result from Eq.~\eqref{eq:MSE3},  
using $r_{\text{inf}}(n) = 0.334 + 0.658(n-1)/(N-1)$,  
is shown as a solid line.    
The semi-logarithmic plot on the left reveals the overall dependence of MSE on $n$,  
whereas the double logarithmic plot on the right highlights the dependence of the negative slope $-2r_{\text{inf}}(n)$ on $n$.  

From the left figure, we observe that the plots for $q < 0.01$ and $0.01 \leq q < 0.1$  
align well with the theoretical predictions.  
Similarly, in the right figure, the slopes of these plots coincide with the theoretical predictions,  
suggesting that the proposed model accurately captures the dynamics of the voting process.  
However, for $0.1 \leq q < 0.4$ and $q \geq 0.4$, the plots decrease more slowly than the theoretical predictions.  
The absolute values of their slopes in the right figure are also smaller than the theoretical values.  

As noted in the previous subsection, the dependence of $r_{\text{inf}}(n)$ on $n$  
for data with $q \geq 0.4$ differs from that for the case of all horses.  
For $n \leq 30$, $r_{\text{inf}}(n)$ remains approximately 0.4, then increases to 1.  
When $r_{\text{inf}}(n) = 0.4$, the power-law exponent for MSE is $2r_{\text{inf}} = 0.8$.  
In the right figure, we plot a straight line with a slope of $-0.5$, which agrees well with 
the plot for $q\geq 0.4$. In order to understand the discrepancy, one needs to 
estimate $r_{\text{inf}}(m)$ more accurately.

\section{\label{sec:conclusion}Conclusion}
In this study, we analyzed the time evolution of single-win odds in Japanese horse racing using a stochastic framework. By modeling the probability dynamics of vote shares, we derived an Ornstein-Uhlenbeck (O-U) process that captures the interaction between bettors who follow market trends (herders) and those who rely on fundamental winning probabilities (informed voter). 
Through empirical analysis of 3,450 Japan Racing Association races in 2008, we demonstrated that the proportion of informed voters increases over time, leading to a more efficient market as the race approaches.

Our results show that the mean squared error (MSE) between the estimated and final vote shares follows a power-law decay, with the exponent dependent on the time evolution of informed bettors. When the proportion of informed bettors increases linearly, the MSE exhibits a crossover from an initial power-law decay to an accelerated exponential decay. This suggests that the betting market undergoes a transition towards efficiency as more bettors incorporate fundamental information, aligning with findings from financial markets where trend-followers and fundamentalists collectively drive price dynamics.

However, our analysis also indicates that for horses with high win probabilities, the transition towards efficiency occurs more slowly. This contributes to the observed favorite-longshot bias, where bettors tend to undervalue favorites and overvalue longshots. Such biases persist despite the overall efficiency gains observed in the market.

Overall, our study highlights the parallels between betting markets and financial markets, demonstrating how simple behavioral interactions lead to non-equilibrium mean-reverting dynamics. This framework provides a robust approach for studying short-lived market behaviors and may be extended to other domains where agents make sequential decisions under uncertainty. Future research could explore extensions incorporating real-time external information sources or applying the model to other wagering systems with varying market structures.

\begin{acknowledgments}
This work was supported by JPSJ KAKENHI [Grant No.{} 22K03445]. 
\end{acknowledgments}

\bibliography{ref.bib}% Produces the bibliography via BibTeX.

\providecommand{\noopsort}[1]{}\providecommand{\singleletter}[1]{#1}%
\begin{thebibliography}{28}%
\makeatletter
\providecommand \@ifxundefined [1]{%
 \@ifx{#1\undefined}
}%
\providecommand \@ifnum [1]{%
 \ifnum #1\expandafter \@firstoftwo
 \else \expandafter \@secondoftwo
 \fi
}%
\providecommand \@ifx [1]{%
 \ifx #1\expandafter \@firstoftwo
 \else \expandafter \@secondoftwo
 \fi
}%
\providecommand \natexlab [1]{#1}%
\providecommand \enquote  [1]{``#1''}%
\providecommand \bibnamefont  [1]{#1}%
\providecommand \bibfnamefont [1]{#1}%
\providecommand \citenamefont [1]{#1}%
\providecommand \href@noop [0]{\@secondoftwo}%
\providecommand \href [0]{\begingroup \@sanitize@url \@href}%
\providecommand \@href[1]{\@@startlink{#1}\@@href}%
\providecommand \@@href[1]{\endgroup#1\@@endlink}%
\providecommand \@sanitize@url [0]{\catcode `\\12\catcode `\$12\catcode
  `\&12\catcode `\#12\catcode `\^12\catcode `\_12\catcode `\%12\relax}%
\providecommand \@@startlink[1]{}%
\providecommand \@@endlink[0]{}%
\providecommand \url  [0]{\begingroup\@sanitize@url \@url }%
\providecommand \@url [1]{\endgroup\@href {#1}{\urlprefix }}%
\providecommand \urlprefix  [0]{URL }%
\providecommand \Eprint [0]{\href }%
\providecommand \doibase [0]{https://doi.org/}%
\providecommand \selectlanguage [0]{\@gobble}%
\providecommand \bibinfo  [0]{\@secondoftwo}%
\providecommand \bibfield  [0]{\@secondoftwo}%
\providecommand \translation [1]{[#1]}%
\providecommand \BibitemOpen [0]{}%
\providecommand \bibitemStop [0]{}%
\providecommand \bibitemNoStop [0]{.\EOS\space}%
\providecommand \EOS [0]{\spacefactor3000\relax}%
\providecommand \BibitemShut  [1]{\csname bibitem#1\endcsname}%
\let\auto@bib@innerbib\@empty
%</preamble>
\bibitem [{\citenamefont {Griffith}(1949)}]{Griffith:1949}%
  \BibitemOpen
  \bibfield  {author} {\bibinfo {author} {\bibfnamefont {R.~M.}\ \bibnamefont
  {Griffith}},\ }\bibfield  {title} {\bibinfo {title} {Odds adjustments by
  american horse-race bettors.},\ }\href {https://doi.org/10.2307/1418275}
  {\bibfield  {journal} {\bibinfo  {journal} {The American journal of
  psychology}\ }\textbf {\bibinfo {volume} {62}},\ \bibinfo {pages} {290}
  (\bibinfo {year} {1949})}\BibitemShut {NoStop}%
\bibitem [{\citenamefont {Ziemba}\ and\ \citenamefont
  {Hausch}(1987)}]{Ziemba:1987}%
  \BibitemOpen
  \bibfield  {author} {\bibinfo {author} {\bibfnamefont {W.~T.}\ \bibnamefont
  {Ziemba}}\ and\ \bibinfo {author} {\bibfnamefont {D.~B.}\ \bibnamefont
  {Hausch}},\ }\href@noop {} {\emph {\bibinfo {title} {Betting at the
  Racetrack}}}\ (\bibinfo  {publisher} {Dr Z Investment Inc.},\ \bibinfo
  {address} {San Luis Obispo, California},\ \bibinfo {year} {1987})\BibitemShut
  {NoStop}%
\bibitem [{\citenamefont {Hausch}\ \emph {et~al.}(2008)\citenamefont {Hausch},
  \citenamefont {Lo},\ and\ \citenamefont {Ziemba}}]{Ziemba:2008}%
  \BibitemOpen
  \bibfield  {author} {\bibinfo {author} {\bibfnamefont {D.~B.}\ \bibnamefont
  {Hausch}}, \bibinfo {author} {\bibfnamefont {V.~S.}\ \bibnamefont {Lo}},\
  and\ \bibinfo {author} {\bibfnamefont {W.~T.}\ \bibnamefont {Ziemba}},\
  }\href {https://doi.org/10.1142/6910} {\emph {\bibinfo {title} {Efficiency of
  Racetrack Betting Markets}}},\ \bibinfo {edition} {2008th}\ ed.\ (\bibinfo
  {publisher} {WORLD SCIENTIFIC},\ \bibinfo {year} {2008})\ \Eprint
  {https://arxiv.org/abs/https://worldscientific.com/doi/pdf/10.1142/6910}
  {https://worldscientific.com/doi/pdf/10.1142/6910} \BibitemShut {NoStop}%
\bibitem [{\citenamefont {Mantegna}\ and\ \citenamefont
  {Stanley}(1999)}]{Mantegna:1999}%
  \BibitemOpen
  \bibfield  {author} {\bibinfo {author} {\bibfnamefont {R.~N.}\ \bibnamefont
  {Mantegna}}\ and\ \bibinfo {author} {\bibfnamefont {H.~E.}\ \bibnamefont
  {Stanley}},\ }\href {https://doi.org/10.1017/CBO9780511755767} {\emph
  {\bibinfo {title} {An Introduction to Econophysics: Correlations and
  Complexity in Finance}}}\ (\bibinfo  {publisher} {Cambridge University
  Press},\ \bibinfo {year} {1999})\BibitemShut {NoStop}%
\bibitem [{\citenamefont {Galam}(2008)}]{Galam:2008}%
  \BibitemOpen
  \bibfield  {author} {\bibinfo {author} {\bibfnamefont {S.}~\bibnamefont
  {Galam}},\ }\bibfield  {title} {\bibinfo {title} {Sociophysics: A review of
  galam models},\ }\href {https://doi.org/10.1142/S0129183108012297} {\bibfield
   {journal} {\bibinfo  {journal} {International Journal of Modern Physics C}\
  }\textbf {\bibinfo {volume} {19}},\ \bibinfo {pages} {409} (\bibinfo {year}
  {2008})},\ \Eprint
  {https://arxiv.org/abs/https://doi.org/10.1142/S0129183108012297}
  {https://doi.org/10.1142/S0129183108012297} \BibitemShut {NoStop}%
\bibitem [{\citenamefont {Galam}(2012)}]{Galam:2012}%
  \BibitemOpen
  \bibfield  {author} {\bibinfo {author} {\bibfnamefont {S.}~\bibnamefont
  {Galam}},\ }\href {https://doi.org/10.1007/978-1-4614-2032-3} {\emph
  {\bibinfo {title} {Sociophysics: A Physicist's Modeling of Psycho-political
  Phenomena}}},\ \bibinfo {edition} {1st}\ ed.,\ Understanding Complex Systems\
  (\bibinfo  {publisher} {Springer New York, NY},\ \bibinfo {year} {2012})\
  pp.\ \bibinfo {pages} {XXIII, 439}\BibitemShut {NoStop}%
\bibitem [{\citenamefont {Park}\ and\ \citenamefont
  {Domany}(2001)}]{Park:2001}%
  \BibitemOpen
  \bibfield  {author} {\bibinfo {author} {\bibfnamefont {K.}~\bibnamefont
  {Park}}\ and\ \bibinfo {author} {\bibfnamefont {E.}~\bibnamefont {Domany}},\
  }\bibfield  {title} {\bibinfo {title} {Power law distribution of dividends in
  horse races},\ }\href {https://doi.org/10.1209/epl/i2001-00169-0} {\bibfield
  {journal} {\bibinfo  {journal} {Europhysics Letters}\ }\textbf {\bibinfo
  {volume} {53}},\ \bibinfo {pages} {419} (\bibinfo {year} {2001})}\BibitemShut
  {NoStop}%
\bibitem [{\citenamefont {Ichinomiya}(2006)}]{Ichinomiya:2006}%
  \BibitemOpen
  \bibfield  {author} {\bibinfo {author} {\bibfnamefont {T.}~\bibnamefont
  {Ichinomiya}},\ }\bibfield  {title} {\bibinfo {title} {Power-law distribution
  in japanese racetrack betting},\ }\href
  {https://doi.org/https://doi.org/10.1016/j.physa.2005.12.027} {\bibfield
  {journal} {\bibinfo  {journal} {Physica A: Statistical Mechanics and its
  Applications}\ }\textbf {\bibinfo {volume} {368}},\ \bibinfo {pages} {207}
  (\bibinfo {year} {2006})}\BibitemShut {NoStop}%
\bibitem [{\citenamefont {Mori}\ and\ \citenamefont
  {Hisakado}(2010{\natexlab{a}})}]{Mori:2010}%
  \BibitemOpen
  \bibfield  {author} {\bibinfo {author} {\bibfnamefont {S.}~\bibnamefont
  {Mori}}\ and\ \bibinfo {author} {\bibfnamefont {M.}~\bibnamefont
  {Hisakado}},\ }\bibfield  {title} {\bibinfo {title} {Exact scale invariance
  in mixing of binary candidates in voting model},\ }\href
  {https://doi.org/10.1143/JPSJ.79.034001} {\bibfield  {journal} {\bibinfo
  {journal} {Journal of the Physical Society of Japan}\ }\textbf {\bibinfo
  {volume} {79}},\ \bibinfo {pages} {034001} (\bibinfo {year}
  {2010}{\natexlab{a}})},\ \Eprint
  {https://arxiv.org/abs/https://doi.org/10.1143/JPSJ.79.034001}
  {https://doi.org/10.1143/JPSJ.79.034001} \BibitemShut {NoStop}%
\bibitem [{\citenamefont {Uhlenbeck}\ and\ \citenamefont
  {Ornstein}(1930)}]{Uhlenbeck:1930}%
  \BibitemOpen
  \bibfield  {author} {\bibinfo {author} {\bibfnamefont {G.~E.}\ \bibnamefont
  {Uhlenbeck}}\ and\ \bibinfo {author} {\bibfnamefont {L.~S.}\ \bibnamefont
  {Ornstein}},\ }\bibfield  {title} {\bibinfo {title} {On the theory of the
  brownian motion},\ }\href {https://doi.org/10.1103/PhysRev.36.823} {\bibfield
   {journal} {\bibinfo  {journal} {Phys. Rev.}\ }\textbf {\bibinfo {volume}
  {36}},\ \bibinfo {pages} {823} (\bibinfo {year} {1930})}\BibitemShut
  {NoStop}%
\bibitem [{\citenamefont {Mori}\ and\ \citenamefont
  {Hisakado}(2010{\natexlab{b}})}]{Mori:2010_2}%
  \BibitemOpen
  \bibfield  {author} {\bibinfo {author} {\bibfnamefont {S.}~\bibnamefont
  {Mori}}\ and\ \bibinfo {author} {\bibfnamefont {M.}~\bibnamefont
  {Hisakado}},\ }\bibfield  {title} {\bibinfo {title} {Component ratios of
  independent and herding betters in a racetrack betting market}} (\bibinfo
  {year} {2010}{\natexlab{b}}),\ \bibinfo {note} {arXiv preprint:
  \href{https://doi.org/10.48550/arXiv.1006.4884}{arXiv:1006.4884}}\BibitemShut
  {NoStop}%
\bibitem [{\citenamefont {Kanazawa}\ \emph {et~al.}(2018)\citenamefont
  {Kanazawa}, \citenamefont {Sueshige}, \citenamefont {Takayasu},\ and\
  \citenamefont {Takayasu}}]{Kanazawa2018}%
  \BibitemOpen
  \bibfield  {author} {\bibinfo {author} {\bibfnamefont {K.}~\bibnamefont
  {Kanazawa}}, \bibinfo {author} {\bibfnamefont {T.}~\bibnamefont {Sueshige}},
  \bibinfo {author} {\bibfnamefont {H.}~\bibnamefont {Takayasu}},\ and\
  \bibinfo {author} {\bibfnamefont {M.}~\bibnamefont {Takayasu}},\ }\bibfield
  {title} {\bibinfo {title} {Derivation of the boltzmann equation for financial
  brownian motion: Direct observation of the collective motion of
  high-frequency traders},\ }\href
  {https://doi.org/10.1103/PhysRevLett.120.138301} {\bibfield  {journal}
  {\bibinfo  {journal} {Physical Review Letters}\ }\textbf {\bibinfo {volume}
  {120}},\ \bibinfo {pages} {138301} (\bibinfo {year} {2018})}\BibitemShut
  {NoStop}%
\bibitem [{\citenamefont {Lux}(1995)}]{Lux1995}%
  \BibitemOpen
  \bibfield  {author} {\bibinfo {author} {\bibfnamefont {T.}~\bibnamefont
  {Lux}},\ }\bibfield  {title} {\bibinfo {title} {Herd behaviour, bubbles and
  crashes},\ }\href {https://doi.org/10.2307/2235156} {\bibfield  {journal}
  {\bibinfo  {journal} {The Economic Journal}\ }\textbf {\bibinfo {volume}
  {105}},\ \bibinfo {pages} {881} (\bibinfo {year} {1995})}\BibitemShut
  {NoStop}%
\bibitem [{\citenamefont {Bouchaud}\ \emph {et~al.}(2002)\citenamefont
  {Bouchaud}, \citenamefont {Mézard},\ and\ \citenamefont
  {Potters}}]{Bouchaud2002}%
  \BibitemOpen
  \bibfield  {author} {\bibinfo {author} {\bibfnamefont {J.-P.}\ \bibnamefont
  {Bouchaud}}, \bibinfo {author} {\bibfnamefont {M.}~\bibnamefont {Mézard}},\
  and\ \bibinfo {author} {\bibfnamefont {M.}~\bibnamefont {Potters}},\
  }\bibfield  {title} {\bibinfo {title} {Statistical properties of stock order
  books: Empirical results and models},\ }\href
  {https://doi.org/10.1088/1469-7688/2/4/301} {\bibfield  {journal} {\bibinfo
  {journal} {Quantitative Finance}\ }\textbf {\bibinfo {volume} {2}},\ \bibinfo
  {pages} {251} (\bibinfo {year} {2002})}\BibitemShut {NoStop}%
\bibitem [{\citenamefont {Alfarano}\ \emph {et~al.}(2005)\citenamefont
  {Alfarano}, \citenamefont {Lux},\ and\ \citenamefont
  {Wagner}}]{Alfarano2005}%
  \BibitemOpen
  \bibfield  {author} {\bibinfo {author} {\bibfnamefont {S.}~\bibnamefont
  {Alfarano}}, \bibinfo {author} {\bibfnamefont {T.}~\bibnamefont {Lux}},\ and\
  \bibinfo {author} {\bibfnamefont {F.}~\bibnamefont {Wagner}},\ }\bibfield
  {title} {\bibinfo {title} {Estimation of agent-based models: The case of an
  asymmetric herding model},\ }\href
  {https://doi.org/10.1007/s10614-005-6415-1} {\bibfield  {journal} {\bibinfo
  {journal} {Computational Economics}\ }\textbf {\bibinfo {volume} {26}},\
  \bibinfo {pages} {19} (\bibinfo {year} {2005})}\BibitemShut {NoStop}%
\bibitem [{\citenamefont {Takayasu}\ \emph {et~al.}(2006)\citenamefont
  {Takayasu}, \citenamefont {Mizuno},\ and\ \citenamefont
  {Takayasu}}]{MTAKAYASU2006}%
  \BibitemOpen
  \bibfield  {author} {\bibinfo {author} {\bibfnamefont {M.}~\bibnamefont
  {Takayasu}}, \bibinfo {author} {\bibfnamefont {T.}~\bibnamefont {Mizuno}},\
  and\ \bibinfo {author} {\bibfnamefont {H.}~\bibnamefont {Takayasu}},\
  }\bibfield  {title} {\bibinfo {title} {Potential force observed in market
  dynamics},\ }\href
  {https://doi.org/https://doi.org/10.1016/j.physa.2006.04.041} {\bibfield
  {journal} {\bibinfo  {journal} {Physica A: Statistical Mechanics and its
  Applications}\ }\textbf {\bibinfo {volume} {370}},\ \bibinfo {pages} {91}
  (\bibinfo {year} {2006})},\ \bibinfo {note} {econophysics
  Colloquium}\BibitemShut {NoStop}%
\bibitem [{\citenamefont {Mizuno}\ \emph {et~al.}(2007)\citenamefont {Mizuno},
  \citenamefont {Takayasu},\ and\ \citenamefont {Takayasu}}]{MIZUNO2007}%
  \BibitemOpen
  \bibfield  {author} {\bibinfo {author} {\bibfnamefont {T.}~\bibnamefont
  {Mizuno}}, \bibinfo {author} {\bibfnamefont {H.}~\bibnamefont {Takayasu}},\
  and\ \bibinfo {author} {\bibfnamefont {M.}~\bibnamefont {Takayasu}},\
  }\bibfield  {title} {\bibinfo {title} {Analysis of price diffusion in
  financial markets using puck model},\ }\href
  {https://doi.org/https://doi.org/10.1016/j.physa.2007.02.049} {\bibfield
  {journal} {\bibinfo  {journal} {Physica A: Statistical Mechanics and its
  Applications}\ }\textbf {\bibinfo {volume} {382}},\ \bibinfo {pages} {187}
  (\bibinfo {year} {2007})},\ \bibinfo {note} {applications of Physics in
  Financial Analysis}\BibitemShut {NoStop}%
\bibitem [{\citenamefont {Takayasu}\ \emph {et~al.}(2007)\citenamefont
  {Takayasu}, \citenamefont {Mizuno},\ and\ \citenamefont
  {Takayasu}}]{MTAKAYASU2007}%
  \BibitemOpen
  \bibfield  {author} {\bibinfo {author} {\bibfnamefont {M.}~\bibnamefont
  {Takayasu}}, \bibinfo {author} {\bibfnamefont {T.}~\bibnamefont {Mizuno}},\
  and\ \bibinfo {author} {\bibfnamefont {H.}~\bibnamefont {Takayasu}},\
  }\bibfield  {title} {\bibinfo {title} {Theoretical analysis of potential
  forces in markets},\ }\href
  {https://doi.org/https://doi.org/10.1016/j.physa.2007.04.094} {\bibfield
  {journal} {\bibinfo  {journal} {Physica A: Statistical Mechanics and its
  Applications}\ }\textbf {\bibinfo {volume} {383}},\ \bibinfo {pages} {115}
  (\bibinfo {year} {2007})},\ \bibinfo {note} {econophysics Colloquium 2006 and
  Third Bonzenfreies Colloquium}\BibitemShut {NoStop}%
\bibitem [{\citenamefont {Yamashita Rios~de Sousa}\ \emph
  {et~al.}(2019)\citenamefont {Yamashita Rios~de Sousa}, \citenamefont
  {Takayasu},\ and\ \citenamefont {Takayasu}}]{Yamashita2019}%
  \BibitemOpen
  \bibfield  {author} {\bibinfo {author} {\bibfnamefont {A.~M.}\ \bibnamefont
  {Yamashita Rios~de Sousa}}, \bibinfo {author} {\bibfnamefont
  {H.}~\bibnamefont {Takayasu}},\ and\ \bibinfo {author} {\bibfnamefont
  {M.}~\bibnamefont {Takayasu}},\ }\bibfield  {title} {\bibinfo {title} {Random
  coefficient autoregressive processes and the puck model with fluctuating
  potential},\ }\href {https://doi.org/10.1088/1742-5468/aaf109} {\bibfield
  {journal} {\bibinfo  {journal} {Journal of Statistical Mechanics: Theory and
  Experiment}\ }\textbf {\bibinfo {volume} {2019}},\ \bibinfo {pages} {013403}
  (\bibinfo {year} {2019})}\BibitemShut {NoStop}%
\bibitem [{\citenamefont {Hisakado}\ and\ \citenamefont
  {Mori}(2010)}]{Hisakado:2010}%
  \BibitemOpen
  \bibfield  {author} {\bibinfo {author} {\bibfnamefont {M.}~\bibnamefont
  {Hisakado}}\ and\ \bibinfo {author} {\bibfnamefont {S.}~\bibnamefont
  {Mori}},\ }\bibfield  {title} {\bibinfo {title} {Phase transition and
  information cascade in a voting model},\ }\href
  {https://doi.org/10.1088/1751-8113/43/31/315207} {\bibfield  {journal}
  {\bibinfo  {journal} {Journal of Physics A: Mathematical and Theoretical}\
  }\textbf {\bibinfo {volume} {43}},\ \bibinfo {pages} {315207} (\bibinfo
  {year} {2010})}\BibitemShut {NoStop}%
\bibitem [{\citenamefont {Mori}\ \emph {et~al.}(2013)\citenamefont {Mori},
  \citenamefont {Hisakado},\ and\ \citenamefont {Takahashi}}]{Mori:2013}%
  \BibitemOpen
  \bibfield  {author} {\bibinfo {author} {\bibfnamefont {S.}~\bibnamefont
  {Mori}}, \bibinfo {author} {\bibfnamefont {M.}~\bibnamefont {Hisakado}},\
  and\ \bibinfo {author} {\bibfnamefont {T.}~\bibnamefont {Takahashi}},\
  }\bibfield  {title} {\bibinfo {title} {Collective adoption of max–min
  strategy in an information cascade voting experiment},\ }\href
  {https://doi.org/10.7566/JPSJ.82.084004} {\bibfield  {journal} {\bibinfo
  {journal} {Journal of the Physical Society of Japan}\ }\textbf {\bibinfo
  {volume} {82}},\ \bibinfo {pages} {084004} (\bibinfo {year} {2013})},\
  \Eprint {https://arxiv.org/abs/https://doi.org/10.7566/JPSJ.82.084004}
  {https://doi.org/10.7566/JPSJ.82.084004} \BibitemShut {NoStop}%
\bibitem [{\citenamefont {Manski}(2006)}]{Manski2006}%
  \BibitemOpen
  \bibfield  {author} {\bibinfo {author} {\bibfnamefont {C.~F.}\ \bibnamefont
  {Manski}},\ }\bibfield  {title} {\bibinfo {title} {Interpreting the
  predictions of prediction markets},\ }\href
  {https://doi.org/10.1016/j.econlet.2006.01.015} {\bibfield  {journal}
  {\bibinfo  {journal} {Economics Letters}\ }\textbf {\bibinfo {volume} {91}},\
  \bibinfo {pages} {425} (\bibinfo {year} {2006})}\BibitemShut {NoStop}%
\bibitem [{\citenamefont {Wolfers}\ and\ \citenamefont
  {Zitzewitz}(2006)}]{Wolfers2006}%
  \BibitemOpen
  \bibfield  {author} {\bibinfo {author} {\bibfnamefont {J.}~\bibnamefont
  {Wolfers}}\ and\ \bibinfo {author} {\bibfnamefont {E.}~\bibnamefont
  {Zitzewitz}},\ }\bibfield  {title} {\bibinfo {title} {Prediction markets in
  theory and practice},\ }\href {https://www.nber.org/papers/w12083} {\bibfield
   {journal} {\bibinfo  {journal} {NBER Working Paper No. 12083}\ } (\bibinfo
  {year} {2006})}\BibitemShut {NoStop}%
\bibitem [{\citenamefont {Gardiner}(2009)}]{Gardiner:2009}%
  \BibitemOpen
  \bibfield  {author} {\bibinfo {author} {\bibfnamefont {C.}~\bibnamefont
  {Gardiner}},\ }\href {https://doi.org/10.1007/978-3-540-70713-4} {\emph
  {\bibinfo {title} {Stochastic Methods: A Handbook for the Natural and Social
  Sciences}}},\ \bibinfo {edition} {4th}\ ed.,\ Springer Series in Synergetics\
  (\bibinfo  {publisher} {Springer Berlin, Heidelberg},\ \bibinfo {year}
  {2009})\ pp.\ \bibinfo {pages} {XVIII, 447}\BibitemShut {NoStop}%
\bibitem [{\citenamefont {Mori}\ and\ \citenamefont
  {Hisakado}(2015)}]{Mori:2015}%
  \BibitemOpen
  \bibfield  {author} {\bibinfo {author} {\bibfnamefont {S.}~\bibnamefont
  {Mori}}\ and\ \bibinfo {author} {\bibfnamefont {M.}~\bibnamefont
  {Hisakado}},\ }\bibfield  {title} {\bibinfo {title} {Finite-size scaling
  analysis of binary stochastic processes and universality classes of
  information cascade phase transition},\ }\href
  {https://doi.org/10.7566/JPSJ.84.054001} {\bibfield  {journal} {\bibinfo
  {journal} {Journal of the Physical Society of Japan}\ }\textbf {\bibinfo
  {volume} {84}},\ \bibinfo {pages} {054001} (\bibinfo {year}
  {2015})}\BibitemShut {NoStop}%
\bibitem [{\citenamefont {Hod}\ and\ \citenamefont {Keshet}(2004)}]{Hod:2004}%
  \BibitemOpen
  \bibfield  {author} {\bibinfo {author} {\bibfnamefont {S.}~\bibnamefont
  {Hod}}\ and\ \bibinfo {author} {\bibfnamefont {U.}~\bibnamefont {Keshet}},\
  }\bibfield  {title} {\bibinfo {title} {Phase transition in random walks with
  long-range correlations},\ }\href
  {https://doi.org/10.1103/PhysRevE.70.015104} {\bibfield  {journal} {\bibinfo
  {journal} {Phys. Rev. E}\ }\textbf {\bibinfo {volume} {70}},\ \bibinfo
  {pages} {015104} (\bibinfo {year} {2004})}\BibitemShut {NoStop}%
\bibitem [{\citenamefont {Huillet}(2008)}]{Huillet:2008}%
  \BibitemOpen
  \bibfield  {author} {\bibinfo {author} {\bibfnamefont {T.}~\bibnamefont
  {Huillet}},\ }\bibfield  {title} {\bibinfo {title} {On
  p$\acute{o}$lya-friedman random walks},\ }\href@noop {} {\bibfield  {journal}
  {\bibinfo  {journal} {J.Phys.A}\ }\textbf {\bibinfo {volume} {41}},\ \bibinfo
  {pages} {505005} (\bibinfo {year} {2008})}\BibitemShut {NoStop}%
\bibitem [{\citenamefont {Sugawara}(2025)}]{DA}%
  \BibitemOpen
  \bibfield  {author} {\bibinfo {author} {\bibfnamefont {T.}~\bibnamefont
  {Sugawara}},\ }\href@noop {} {\bibinfo {title} {Time series analysis of 2008
  jra win odds}},\ \bibinfo {howpublished}
  {\url{https://github.com/LABO-M/Ornstein-Uhlenbeck-Process-for-Horse-Race-Betting}}
  (\bibinfo {year} {2025}),\ \bibinfo {note} {gitHub repository}\BibitemShut
  {NoStop}%
\end{thebibliography}%

\appendix

\section{Derivation of eq.\eqref{eq:SDE}}
We start from the difference equation for $Z(n,q)$,
\begin{eqnarray}
\Delta Z(n,q)&=&Z(n+1,q)-Z(n,q)=
\frac{1}{(n+1)\Delta t}\sum_{s=1}^{(n+1)\Delta t}X(s,q)
-\frac{1}{n\Delta t}\sum_{s=1}^{n\Delta t}X(s,q) \nonumber \\
&=& \frac{1}{n+1}\left(\frac{1}{\Delta t}\sum_{s=n\Delta t+1}^{(n+1)\Delta t}X(s,q)-Z(n,q)\right)  \nonumber.
\end{eqnarray}
The conditional expected value of $\Delta Z(n,t)$ under $Z(n,t)=z$ is,
\[
\mathbb{E}[\Delta Z(n,q)|Z(n,q)=z]=\frac{1}{n+1}(f_n(z)-z)=
-\frac{1}{n+1}r_{\text{inf}}(n)(z-q) \label{eq:micro2}.
\]
The voters during the $n$-th interval
 vote conditionally independent and the variance of $\Delta Z(n,t)$ is 
 the sum of the variance of $\{X(t,q)\}$ in the interval.
The conditional variance of $\Delta Z(n,t)$ under $Z(n,t)=z$ is,
\[
\mathbb{V}(\Delta Z(n,q)|Z(n,q)=z)=\frac{1}{(n+1)^2\Delta t}f_n(z)(1-f_n(z)).
\]
The conditional expected value and the square root of the 
conditional variance of $Z(n,q)$ give the drift term and the
fluctuation term of the stochastic differential equation 
of $Z(n,t)$ as\cite{Gardiner:2009},
\[
dZ(n,q)=\mathbb{E}[\Delta Z(n,q)|Z(n,q)=z]dn+\sqrt{\mathbb{V}(\Delta Z(n,q)|Z(n,q)=z)}dW(n,q).
\]
As we have shown $r_{\text{inf}}(n)>0$ in Section \ref{sec:data}, the process is mean-reverting 
and the MSE between $Z(n,q)$ and $q$ converges to zero.
We can assume that $Z(n,q)\simeq q$ and  simplify $f_{n}(z)(1-f_n(z))$ as $q(1-q)$ for large $n$.
The stochastic differential equation of $Z(n,q)$ is written as, 
\[
dZ(n,q)=-\frac{r_{\text{inf}}(n)(Z(n,q)-q)}{n+1}dn+\frac{\sqrt{q(1-q)}}{(n+1)\sqrt{\Delta t}}dW(n,q).
\]

\section{Derivation of eq.\eqref{eq:sol}}
We solve \eqref{eq:SDE} and obtain the solution of 
eq.(\ref{eq:sol}). We rewrite \eqref{eq:SDE}
for $Z(n,q)-q$ as,
\[
d(Z(n,q)-q)=-\frac{1}{n+1}r_{\text{inf}}(n)(Z(n,q)-q)dn+\frac{b(q)}{n+1}dW(n,q).
\]
Here, $b(q)=\sqrt{q(1-q)}/\Delta t$.
We assume the next form of the solution and put it 
into the above equation.
\[
Z(n,q)-q=\exp\left(-\int_{1}^{n}\frac{r_{\text{inf}}(m)}{m+1}dm\right)
y(n,q).
\]
$y(n,q)$ satisfies the next SDE,
\[
dy(n,q)=\exp\left(\int_{1}^{n}\frac{r_{\text{inf}}(m)}{m+1}dm\right)\frac{b(q)}{n+1}dW(n,q)
\]
The initial condition of $y(n,q)$ is,
\[
y(1,q)=z(1,q)-q.
\]
We perform the Wiener integral and obtain,
\[
y(n,q)=y(1,q)+\int_{1}^{n}\exp\left(\int_{1}^{l}\frac{r_{\text{inf}}(m)}{m+1}dm\right)\frac{b(q)}{l+1}dW(l,q).
\]
We convert from $y(n,q)$ to $Z(n,q)$ and obtain,
\begin{eqnarray}
Z(n,q)&=&q+e^{-\int_{1}^{n}\frac{r_{\text{inf}}(m)}{m+1}dm}
\left((z(1,q)-q)+
\int_{1}^{n}e^{\int_{1}^{l}\frac{r_{\text{inf}}(m)}{m+1}dm}\left(\frac{b(q)}{l+1}\right)dW(l,q)\right)
\nonumber \\
&=&q+e^{-\int_{1}^{n}\frac{r_{\text{inf}}(m)}{m+1}dm}(z(1,q)-q)+
\int_{1}^{n}e^{-\int_{l}^{n}\frac{r_{\text{inf}}(m)}{m+1}dm}\left(\frac{b(q)}{l+1}\right)dW(l,q).\nonumber 
\end{eqnarray}

\end{document}